\documentclass[onecolumn]{article}
\usepackage[english]{babel}

\usepackage[utf8]{inputenc}
\usepackage{authblk}
\usepackage{dcolumn}
\usepackage{amsmath}
\usepackage{longtable}
\usepackage{tabularx}
\usepackage{setspace}
\usepackage{multirow}
\usepackage{mathtools}
\usepackage{tikz}
\usepackage{lscape}
\usepackage{amssymb}
\usepackage{url}
\usepackage{arydshln}
\usepackage[misc]{ifsym}
\usepackage{apacite}
\usepackage[misc]{ifsym}
\usepackage{booktabs}

\usepackage{graphicx}
\usepackage{subfig}
\usepackage{lipsum}

\RequirePackage[twoside,letterpaper,includeheadfoot,layoutsize={8.125in,10.875in},layouthoffset=0.1875in,layoutvoffset=0.0625in,left=38.5pt,right=43pt,top=43pt,bottom=32pt,headheight=0pt,headsep=10pt,footskip=25pt]{geometry}
\title{Social networks and group configuration in activities at school level}
\title{Randomly Assigned Groups in Secondary Education: Greater Collaboration but Lower Stability}
\title{Affinity-Based Groups in Secondary Education: Increased Stability at the Expense of Collaboration}

\author{
Diego Ram\'irez$^{1,3,*}$, Eugenio J. Guzmán-Lavín$^{1}$, Javier Pulgar$^{2,3}$, Cristian Candia$^{4,5,1}$\\ 
\small{$^{1}$Centro de Investigación en Complejidad Social (CICS), Facultad de Gobierno, Universidad del Desarrollo, Chile.}\\
\small{$^{2}$Departamento de F\'isica, Universidad del B\'io B\'io, Concepci\'on, 4051381, Chile.}\\
\small{$^{3}$Grupo de Investigaci\'{o}n en Did\'{a}ctica de las Ciencias y Matem\'{a}tica (GIDICMA)}\\
\small{$^{4}$Data Science Institute, Facultad de Ingeniería, Universidad del Desarrollo, Las Condes, 7610658, Chile.}\\
\small{$^{5}$Northwestern Institute on Complex Systems (NICO), Northwestern University, Evanston, IL 60208.}\\
\small{$^{*}$Corresponding Author: dieramirez@udd.cl}
}

\begin{document}
\maketitle 
\begin{abstract}

Group configurations play a pivotal role in shaping social skills and learning outcomes by fostering student interaction and collaboration. Yet, the factors influencing group formation, social cohesion, and group stability remain understudied. Here, we examine these factors in a cohort of 90 students from 9th and 10th grades at a private school in Chile. Students were either randomly assigned to groups (control) or allowed to choose their own groups (experimental) for a semester-long project. Our results show that self-selected groups had stronger friendship ties by the end of the semester. In contrast, randomly assigned groups exhibited higher levels of cooperation and more frequent membership changes, indicating weaker social cohesion. Additionally, 10th-grade students were more likely to stay in their original groups, possibly suggesting increased social maturity. These findings underscore the ongoing challenge in education of balancing student choice in group formation with the need for diverse group compositions that enhance learning. The study provides actionable insights for educators.

\end{abstract}

\section{Introduction}

The 21st-century skills framework emphasizes the importance of collaboration, problem-solving, and emotional intelligence alongside traditional academic knowledge. These skills are not developed in isolation but are honed through interactive and collaborative learning experiences. Group work in educational settings serves as a critical platform for students to practice and internalize these skills \cite{borgattiNetworkAnalysisSocial2009a,rambaranAcademicFunctioningPeer2017,vygotskyMindSocietyDevelopment1978a,heinimakiStudentParticipatoryRole2021a}. However, despite its importance, there is a gap in our understanding of how different methods of group formation impact the effectiveness of such collaborative learning.\\

In the classroom, students are commonly grouped in different ways: randomly \cite{chapmanCanWePick2006}, by self-selection (by students' own choice)\cite{connerleyImportanceInvasivenessStudent2001}, through teacher guidance \cite{blowersUsingStudentSkill2003}, or a mix of these. The stability of these groups is a key measure of their effectiveness \cite{hackman_2010}. Stable groups enhance student motivation and commitment by fostering a sense of community \cite{ostermanStudentsNeedBelonging2000,vollingerVideobasedStudyStudent2022,johnsonActiveLearningCooperation2008}. They also allow students to develop important skills like conflict resolution and effective communication, which are valuable in both academic and life settings \cite{johnsonActiveLearningCooperation2008,johnsonCooperativeLearning21st2014}. In stable groups, students are more likely to share ideas and take academic risks, promoting deeper learning and critical thinking \cite{barronWhenSmartGroups2003,gilliesEffectsCooperativeLearning2004,wegerifDialogicEducationTechnology2007,mercerTalkDevelopmentReasoning2008}. However, not all groups achieve stability and effectiveness, making it imperative for educators to closely monitor group dynamics and intervene when necessary \cite{johnsonCooperativeLearning21st2014}. The need for stable and effective groups is especially pressing in today's educational landscape, which highly values social cooperation and collaborative learning. \cite{gilliesBehaviorsInteractionsPerceptions2003,gilliesEffectsCooperativeLearning2004,heinimakiStudentParticipatoryRole2021a}.\\

In this study, we aim to fill this knowledge gap by exploring the dynamics of group membership and change among 9th and 10th-grade students, focusing on their impact on the development of collaboration networks, one of the key 21st-century skills. We compare self-selected groups with randomly assigned ones, examining their impact on social networks and academic performance. Our findings reveal that self-selected groups had stronger friendship ties, while randomly assigned groups showed higher levels of cooperation but less stability. Our findings offer valuable insights for educators striving to prepare students for the collaborative challenges of the future.

\section{Literature Review}

The literature shows that positive group dynamics enhance collaboration, increase engagement, and sharper problem-solving skills \cite{louSmallGroupIndividual2001a,springerEffectsSmallGroupLearning1999}. These dynamics also help develop social and emotional skills like effective communication, empathy, and conflict resolution \cite{moradiRelationshipGroupLearning2018,laybournMeasuringChangesGroupworking2001}. Group activities not only hone these skills but also build social bonds and social competence \cite{zinsBuildingAcademicSuccess2004,lopesRoleKnowledgeSkills2012}. Notably, students who engage in effective group work often achieve higher academic performance, emotional well-being, and show lasting positive outcomes \cite{eliasPromotingSocialEmotional1997,battistichInteractionProcessesStudent1993,candia2022reciprocity,landaeta2023game,candia2022interconnectedness}.\\

Understanding the elements that influence team cohesion is crucial. Team cohesion unites members, minimizes differences, and strengthens group identity \cite{parkEffectCohesionCurvilinear2017}. This unity improves communication, enforces norms, and helps the team respond collectively to pressures for conformity. Cohesion is vital for effective group function and is influenced by several factors. Strong friendships often form in cohesive teams, enhancing trust and communication \cite{hanBridgeBondDiverse2013,herbisonIntricaciesFriendshipcohesionRelationship2017}. This unity also fosters collaboration, creating a shared sense of purpose and mutual effort toward goals \cite{liangTeamDiversityTeam2015,goswamiExplicatingInfluenceShared2020}. Additionally, recognizing individual expertise is key \cite{festingerInformalSocialCommunication1950}, as it allows teams to leverage specialized skills and makes members feel valued for their unique contributions \cite{carmeliLinkingPerceivedExternal2011a}.\\

Modern teaching methods emphasize group activities, linking them to improved critical thinking, increased creativity, and stronger social ties \cite{johnsonActiveLearningCooperation2008,springerEffectsSmallGroupLearning1999}. Therefore, understanding group structures and the reasons for team selection can significantly enhance the educational experience for students \cite{cohenRestructuringClassroomConditions1994,gilliesBehaviorsInteractionsPerceptions2003}.\\

Students' attributes like academic ability, learning styles, and interests significantly influence group formation \cite{getzelsClassroomGroupUnique1960,holmesLeaderFollowerIsolate1980}. Recent studies explore the link between group structure, social networks, and learning outcomes \cite{gasevicChooseYourClassmates2013, candia2022reciprocity, candia2022interconnectedness}. Heterogeneous groups, with diverse skills and backgrounds, often foster robust interactions and idea exchange \cite{wichmannGroupFormationSmallGroup2016,brownMakingGroupBrainstorming2002}. In contrast, homogeneous groups may limit diverse viewpoints, narrowing the scope of learning \cite{larsonDeepDiversityStrong2007}.\\

Group formation methods, whether random, teacher-guided, or self-selected, have drawbacks. Randomly formed groups can be unpredictable, leading to skill imbalances and reduced team effectiveness \cite{chapmanCanWePick2006,johnsonActiveLearningCooperation2008}. Self-selected groups may lean toward homogeneity, limiting perspective diversity and causing feelings of exclusion \cite{srbaDynamicGroupFormation2015a,deibelTeamFormationMethods2005,oakleyTurningStudentGroups}. Teacher-guided groups aim to balance these limitations by leveraging student strengths and weaknesses. However, this approach requires deep pedagogical knowledge and accurate understanding of student dynamics \cite{michaelsenTeambasedLearningTransformative2002}. A mismatch between teacher perceptions and student abilities can still lead to ineffective groups \cite{urhahneReviewAccuracyTeacher2021,wangSystematicReviewTeacher2018}.\\

Common methods for group assignment in educational settings include self-selection and random assignment \cite{chapmanCanWePick2006,wentzelPeerRelationshipsMotivation2017,hanhamGroupWorkSchools2009,mcfarlandNetworkEcologyAdolescent2014}. Research offers mixed views on the primary influences on group dynamics. Some studies suggest that instructor-organized or mixed-method groups lead to better classroom experiences \cite{fiechtnerRepublicationWhyGroups2016a,cestonePeerAssessmentEvaluation2008}. However, Bacon et al. \citeyear{baconLessonsBestWorst1999} found that self-selected groups provided the best team experiences. Despite these differing views, the goal remains the same: to create an optimal group environment that encourages efficient learning and diverse thinking \cite{gasevicChooseYourClassmates2013,choSocialNetworksCommunication2007b}. Educators are thus challenged to balance student autonomy in group selection with the need for diverse representation.\\

Building on this challenge, this study aims to deepen our understanding of group dynamics and stability in educational settings. It examines factors such as friendship-based social preferences, academic reputation, and cooperative tendencies. The research addresses two main questions:

\begin{enumerate}

    \item Do differences exist in team friendship, cooperation, and recognition of academic prestige between students in randomly assigned groups and those in self-selected groups after a semester?
    \item Are there variations in the number of team members with whom a student chooses to continue working, between those in randomly assigned groups and those in self-selected groups, after a semester?
    
\end{enumerate}

\section{Data and Methods}

\subsection{Research Context and Participants}

The study took place in the 2021 academic year at a private school in Chile. Participants were 9th and 10th-grade students (ages 14-16) from two parallel courses. The study focused on physics classes taught by the same teacher. For clarity, the 9th-grade courses are labeled C1 and E1, and the 10th-grade courses as C2 and E2. Active learning methods were used in all four courses. These methods were tailored to each grade level to cover the physics curriculum. Students worked in groups for 10 sessions per semester. Activities included discussion, poster creation, and designing experiments to demonstrate physical concepts.\\

In 2021, as COVID-19 cases decreased and vaccination rates increased, schools began to offer in-person learning. Some students returned to the classroom, while others continued online. The teacher reported that group activities proceeded smoothly in both formats. All students had the means to participate online, either at home or at school, ensuring constant peer contact. By the end of the year, most students had attended in-person sessions.

\begin{table}[!h]
	\centering
	\caption{School Courses information}
	\label{tab:cantidadestudintes}
	{
		\begin{tabular}{cccccc}
			\toprule
			Course & Condition & Grade Level & Female & Male & Total\\
			\cmidrule[0.4pt]{1-6}
                C1 & Control & 9th &$14$ & $10$ & $24$  \\
			C2 & Control & 10th &$11$ & $9$ & $20$  \\
			E1 & Experimental & 9th&$14$ & $10$ & $24$  \\
			E2 & Experimental & 10th&$9$ & $13$ & $22$  \\
               
   \cline{1-6}
   \\
			Total $ $ &&& $48$ & $42$ & $90$  \\
			\bottomrule
		\end{tabular}
	}
\end{table}

\subsection{Data Collection and Intervention}

\begin{figure}[!htbp]
\centering
  \includegraphics[width=0.7\textwidth]{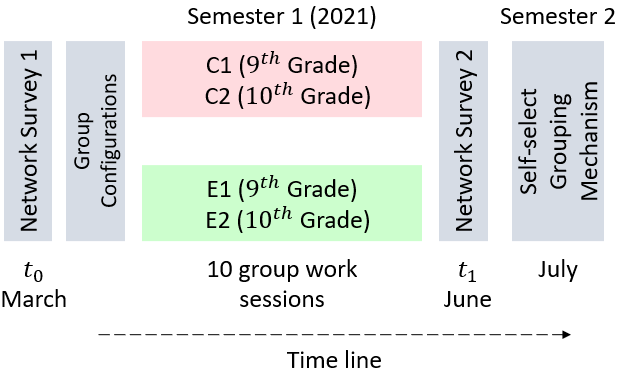}
  \caption{Diagram of the process adopted in the research. Before forming the groups ($t_0$), the first network survey was conducted, which exclusively addressed friendship relationships among students. After the teamwork sessions ($t_1$), the instrument was reapplied and two items related to perceptions of collaboration were incorporated, as well as recognition of prestige in the subject. \label{fig:esquema}}
\end{figure}

The study received approval from the school's institutional committee, and participation was voluntary. Data was collected during the first and second semesters of 2021, from March to July, in line with Chile's academic calendar (see Figure \ref{fig:esquema}). Table \ref{tab:cantidadestudintes} summarizes the total number of participants for each course. We collected data on students' group membership, grades in group activities, and gender. A social networks survey was administered at the start of the first semester (time \( t_0 \)) to measure friendship levels among students \cite{pulgar_etal_2022}. The survey was repeated at the end of the semester (time \( t_1 \)) to include questions on physics prestige and collaboration (see Subsection \ref{sec:instruments}).\\

For the intervention, students in control courses (C1 and C2) were randomly assigned to groups by the teacher at the start of the first semester. Students in experimental courses (E1 and E2) chose their groups. At the start of the second semester, all students had the option to change groups. Six work groups were formed in each course (see Table \ref{tab:descripgrup}). Group composition remained stable during the first semester and could be changed at the start of the second semester (see Figure \ref{fig:esquema}). Students were informed that their grades would be based on group activities and were given specific assessment instructions and rubrics. Each semester included two group activities, each lasting five class sessions, totaling 10 sessions per semester.\\

At the end of the first semester (time \( t_1 \)), a network survey was administered to assess social relationships, including friendship, prestige, and collaboration in physics. At the start of the second semester, students had the option to change their group configurations. The second semester followed the same structure as the first, with students participating in group activities based on their new or existing group configurations.

\subsection{Instruments}
\label{sec:instruments}

We use a trustworthy method for network mapping--an online network survey--to gauge student collaboration \cite{pulgar_etal_2022, pulgar_etal2020}. This survey was created to collect information on different social areas, including friendships, physics-related prestige, and collaborative ties. The survey's specific questions were:

\begin{enumerate}

\item From the students in the classroom roster, indicate those who are your friends.
\item From the students in the classroom roster, indicate those you consider high-achieving physics students.
\item From the students in the classroom roster, indicate those you collaborated with during the classroom activities.

\end{enumerate}


\section{Results}

\subsection{Data Analysis}

\begin{figure}[!htbp]
\centering
  \includegraphics[width=0.6\textwidth]{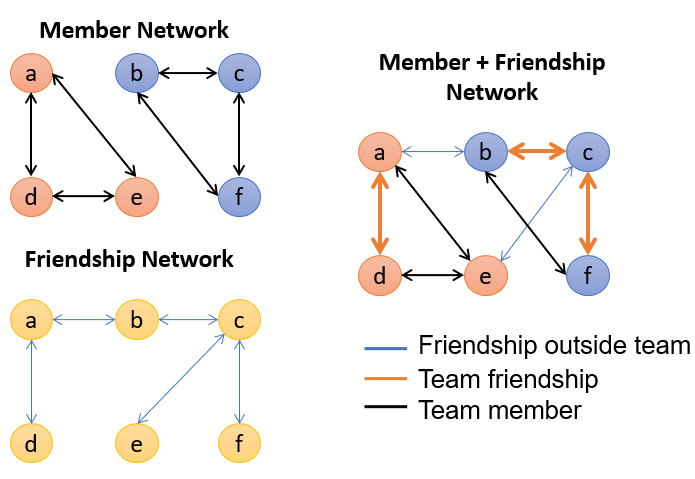}
  \caption{This diagram illustrates how we identify key factors within and between student work groups. Black lines show who belongs to each group, with two example sets of members highlighted in different colors. The friendship network is also shown, where lines connect students who consider each other friends. When we combine these two networks, we get a comprehensive view of friendships both inside (in orange) and outside (in sky blue) of each group. The same approach is used to analyze prestige and cooperation among students. Note that the black lines serve only to show group membership and don't add extra information. \label{fig:metodored}}
\end{figure}

We analyzed student-formed groups using network surveys that measured friendship, prestige, and collaboration. These surveys created indirect networks, capturing relationships that were not necessarily mutual. We also created a membership network to identify each student's specific group affiliation. This membership network was then combined with the indirect networks. This approach allowed us to examine perceptions of friendship, prestige, and collaboration both within and between groups, as shown in Figure \ref{fig:metodored}.

We used Gephi 0.9 software to analyze the networks and measure key characteristics, such as the centrality of each student in the network \cite{ICWSM09154}. These measures include degree, in-degree (how others perceive an individual), and out-degree (how an individual perceives others). These details are in the supplementary material Table \ref{tab:parametrosred2}.\\

For data analysis, we used standard R libraries \cite{rstudio}. We created linear regression models to answer our research questions. To address the first question, we used different models to predict friendship, cooperation, and prestige within each work team after completing group activities (see Table \ref{tab:r2}). For the second research question, we aimed to predict how many team members a student would continue working with after being given the option to change groups. We used a variable that took values from 0 to 3, representing the number of original team members a student continued to work with. We also used the in-degree from the friendship network as a control variable.\\

To ensure the reliability of our findings, we conducted a bootstrapping analysis. This showed no significant changes in the regression coefficients at the 0.01 level. We followed the guidelines of Dou and Zwolak for this analysis \cite{douPractitionerGuideSocial2019}. The bootstrapping results are included in the supplementary material. Additionally, we found no issues of multicollinearity in our models, as confirmed in Tables \ref{tab:s3}, \ref{tab:s4} in the supplementary material. 

\subsection{Social Networks}

We used regression models to analyze team-level network characteristics at the end of the first semester (Table \ref{tab:r2}). These characteristics include in-degree of team friendship (Models 1-4), in-degree of team cooperation (Models 5-6), and in-degree of team prestige (Models 7-8). We controlled for variables like group size, gender, grade level, and average grades. We also used initial team friendship and overall friendship before the intervention as predictors.\\

Model 1 shows that initial team friendship is the only significant predictor (p $<$ 0.001) for team friendship at the end of the semester, explaining $71\%$ of the variance. In Model 2, adding an interaction term improved the model's explanatory power to $75\%$. Models 3 and 4 replace team friendship with initial friendship and find it to be the main predictor (p $<$ 0.001 and p $<$ 0.01). Grade level also emerges as a significant predictor (p $<$ 0.05), but these models explain less variance (52$\%$ and 53$\%$).

\begin{table}[h] \centering 
\footnotesize	
  \caption{Regression models: Predicting within-group friendship, prestige, and cooperation links at time $t_1$} 
  \label{tab:r2} 
\begin{tabular}{lcccc|cccc} 
\cline{1-9}\\
 & \multicolumn{8}{c}{\textit{Dependent variable:}} \\ 
[0.5ex]\cline{2-9}
\\ &  \multicolumn{4}{c}{Team Friendship ($t_1$)} & \multicolumn{2}{c}{ Team Cooperation ($t_1$)} & \multicolumn{2}{c}{Team Prestige ($t_1$) }\\
\\ &  \multicolumn{4}{c}{(in-degree)} & \multicolumn{2}{c}{(in-degree)} & \multicolumn{2}{c}{(in-degree) }\\
\\ 
[0.5ex]\cline{2-9}
	\\&(1)&(2)&(3)&(4)&(5)&(6)&(7)&(8)	\\
 \hline \\ 

Exp. (Affinitty-based groups)	&	0.25	&	0.82$^{**}$	&	1.17$^{***}$	&	1.39$^{***}$	&	-0.70$^{***}$	&	-0.99$^{***}$	&	-0.25	&	-0.04	\\
  	&	(0.18)	&	(0.25)	&	(0.18)	&	(0.38)	&	(0.20)	&	(0.28)	&	(0.26)	&	(0.38)	\\
 Group Size (3)     	&	-0.33	&	-0.39$^{*}$	&	-0.49+	&	-0.49+	&	-1.32$^{***}$	&	-1.29$^{***}$	&	-0.36	&	-0.38	\\
                    	&	(0.20)	&	(0.19)	&	(0.26)	&	(0.26)	&	(0.22)	&	(0.22)	&	(0.29)	&	(0.29)	\\
 Gender (Female)    	&	0.14	&	0.14	&	-0.03	&	-0.02	&	-0.25+	&	-0.25+	&	-0.43$^{*}$	&	-0.43$^{*}$	\\
                    	&	(0.13)	&	(0.13)	&	(0.17)	&	(0.17)	&	(0.14)	&	(0.14)	&	(0.19)	&	(0.19)	\\
  Grade Level (10th)	&	0.06	&	0.08	&	0.46$^{*}$	&	0.45$^{*}$	&	0.89$^{***}$	&	0.88$^{***}$	&	-0.11	&	-0.10	\\
 	&	(0.17)	&	(0.16)	&	(0.22)	&	(0.22)	&	(0.19)	&	(0.19)	&	(0.25)	&	(0.25)	\\
  Average grades    	&	0.02	&	0.02+	&	0.01	&	0.01	&	-0.03$^{*}$	&	-0.03$^{*}$	&	0.05$^{**}$	&	0.05$^{***}$	\\
	&	(0.01)	&	(0.01)	&	(0.01)	&	(0.01)	&	(0.01)	&	(0.01)	&	(0.02)	&	(0.02)	\\
Team Friendship ($t_0$)	&	0.75$^{***}$	&	1.11$^{***}$	&		&		&	0.15	&	-0.03	&	0.34$^{**}$	&	0.47$^{*}$	\\
(in-degree)	&	(0.08)	&	(0.14)	&		&		&	(0.09)	&	(0.16)	&	(0.12)	&	(0.21)	\\
Exp.$*$ Team Friendship ($t_0$)	&		&	-0.54$^{**}$	&		&		&		&	0.27	&		&	-0.20	\\
(in-degree)	&		&	(0.17)	&		&		&		&	(0.19)	&		&	(0.25)	\\
Friendship ($t_0$)	&		&		&	0.12$^{***}$	&	0.15$^{**}$	&		&		&		&		\\
(in-degree)	&		&		&	(0.03)	&	(0.06)	&		&		&		&		\\
Exp.$*$ Friendship ($t_0$)	&		&		&		&	-0.03	&		&		&		&		\\
(in-degree)	&		&		&		&	(0.07)	&		&		&		&		\\
Intercept	&	-0.73	&	-1.08+	&	-0.68	&	-0.80	&	4.13$^{***}$	&	4.30$^{***}$	&	-2.36$^{*}$	&	-2.49$^{*}$	\\
	&	(0.67)	&	(0.64)	&	(0.86)	&	(0.90)	&	(0.72)	&	(0.73)	&	(0.96)	&	(0.98)	\\
	&		&		&		&		&		&		&		&		\\

 \hline \\ 	
Num.Obs.	&	90	    &	90	    &	90	    &	90	&	90	&	90	&	90	&	90	\\
R2	        &	0.71	&	0.75	&	0.52	&	0.53	&	0.39	&	0.40	&	0.28	&	0.29	\\
R2 Adj. 	&	0.69	&	0.72	&	0.49	&	0.49	&	0.34	&	0.35	&	0.23	&	0.23	\\
AIC	        &	178.3	&	169.4	&	223.9	&	225.4	&	193.4	&	193.3	&	244.5	&	245.8	\\
BIC	        &	198.3	&	191.9	&	243.9	&	247.9	&	213.4	&	215.8	&	264.5	&	268.3	\\
Log.Lik.	&	-81.16	&	-75.70	&	-103.93	&	-103.69	&	-88.72	&	-87.63	&	-114.24	&	-113.91	\\
F	        &	34.42	&	34.42	&	15.26	&	13.04	&	8.68	&	7.82	&	5.42	&	4.71	\\
RMSE	    &	0.62	&	0.59	&	0.80	&	0.81	&	0.68	&	0.67	&	0.90	&	0.90	\\
\\
\hline 

\textit{Note:}  & \multicolumn{8}{r}{$^{\dagger}$ p $<$ 0.1, $^{*}$ p $<$ 0.05, $^{**}$ p $<$ 0.01, $^{***}$p $<$ 0.001} \\ 
\end{tabular} 
\end{table} 

For team cooperation (Models 5-6), treatment, group size, and average grade negatively affect cooperation (p $<$ 0.001 for treatment and group size, p $<$ 0.05 for average grades), while grade level has a positive effect (p $<$ 0.001). These models explain $39\%$  and $40\%$  of the variance. The results suggest that control groups, despite having fewer friendships, engaged in more cooperative ties. For prestige (Models 7-8), gender negatively affects prestige (p $<$ 0.05), while average grade and team friendship have a positive effect (p $<$ 0.001 for average grade, p $<$ 0.05 for team friendship). These models explain $28\%$  and $29\%$  of the variance. Gender, average grade, and team friendship are the main predictors, regardless of the treatment condition.

\subsection{Group Changes}

Six work groups were formed in 9th and 10th-grade classrooms in the first semester. In the 9th grade, each group had four members, while in the 10th grade, group sizes varied from 3 to 4 members (Table \ref{tab:descripgrup}). In the second semester, 9th-grade groups remained at four members each, but 10th-grade groups varied in size from two to four members. Regarding gender composition, experimental groups in both grades showed a tendency for same-gender grouping: four all-female and five all-male groups in the first semester. Only three groups had a mix of genders. This pattern continued into the second semester. In contrast, control groups had only one all-female group in the first semester. However, when students had the chance to change groups in the second semester, the number of single-gender groups increased to four all-female and three all-male groups across both grades. 

\begin{table*}[!h]
\begin{center}
\footnotesize	
  \caption{Description of the members of the work teams, specifying the number of members and the percentage of women in each group, both for the first and second semester. \label{tab:descripgrup}}
  \centering
 \begin{tabular}{lccccccc}
\hline
\\[-1.2mm]
&  \multicolumn{3}{c}{Group C1}& & \multicolumn{3}{c}{Group C2}\\
\cline{2-7}
\\[0.1mm]
&  \multicolumn{2}{c}{First semester
}& Second semester &  \multicolumn{2}{c}{First semester}&  \multicolumn{2}{c}{Second semester
}\\[0.5mm]
\cline{2-8}
\\[0.3mm]
Group & Members & \% Women & \% Women & Members & \% Women & Members & \% Women \\
\\[0.3mm]
\hline
\\[0.3mm]

1	&	4	&	50	&	75	&	3	&	     33.3     	&	2	&	100\\
2	&	4	&	25	&	50	&	3	&	   33.3     	&	3	&	0 \\
3	&	4	&	40	&	100	&	3	&	     66.7     	&	3	&	33.3 \\
4	&	4	&	50	&	25	&	3	&	   66.7    &	4	&	0  \\	
5	&	4	&	75	&	0	&	4	&	   50       	&	4	&	100 \\
6	&	4	&	100	&	100	&	4	&	   75       	&	4	&	75 \\

\hline
\\[-1.2mm]
&  \multicolumn{3}{c}{Group E1}&  \multicolumn{4}{c}{Group E2}\\
\\[-1.2mm]
\hline 
\\[-1.2mm]
1	&	4	&	75	&	100	&	3	&	0	&	4	&	100 \\
2	&	4	&	0	&	0	&	4	&	50	&	3	&	100\\
3	&	4	&	100	&	75	&	4	&	100	&	3	&	66.7\\
4	&	4	&	0	&	50	&	3	&	100	&	4	&	0\\
5	&	4	&	75	&	100	&	4	&	0	&	4	&	0\\
6	&	4	&	100	&	25	&	4	&	0	&	4	&	0\\

\hline
\end{tabular}
\end{center}
\end{table*}

\begin{figure}[!h]
\centering
  \includegraphics[width=0.9\textwidth]{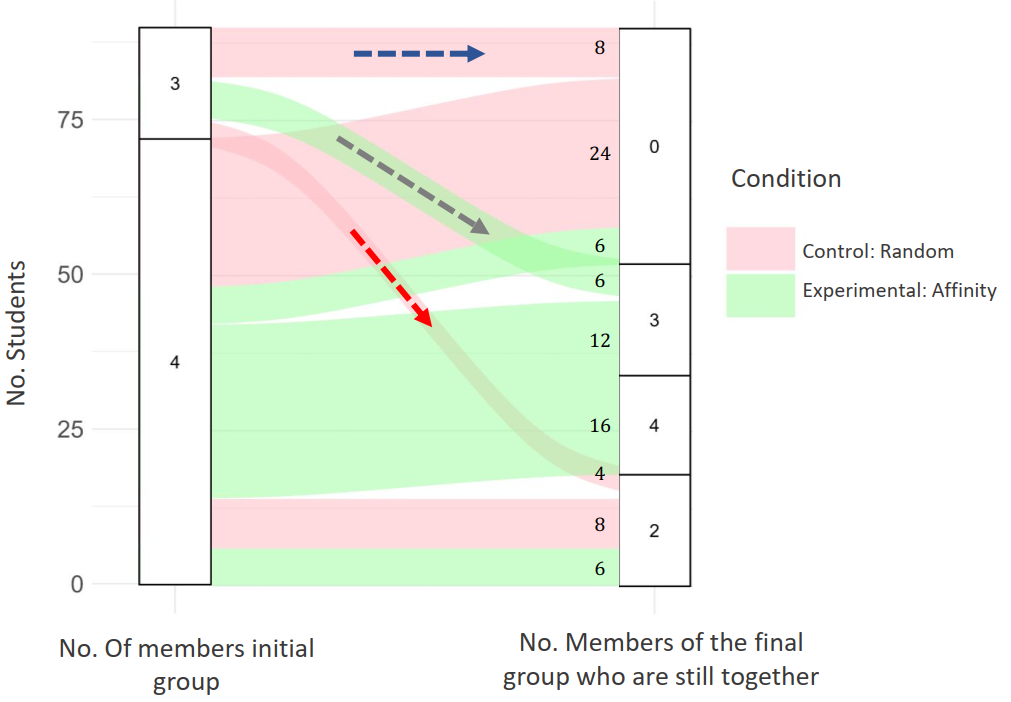}
  \caption{Student group transitions in control and experimental settings. In the control condition, students are randomly assigned to groups (shown in pink), while in the experimental condition, they form groups based on affinity (indicated in green). The arrows, sized proportionally, show the number of students who chose to either stay with their original group members or switch to new peers when given the option. \label{fig:flujo1}}
\end{figure}

To provide a detailed view of group changes, Figure \ref{fig:flujo1} shows the student flow in both experimental and control classrooms across the first and second semesters. Out of 90 students, 38 ended up in new groups without any of their original members; 32 of these ($84\%$) were from control (randomly assigned groups) classrooms C1 and C2. On the other hand, 52 students remained with at least one original group member in the second semester, with 40 of these ($76.9\%$) being from the experimental condition (affinity-based groups).\\

The flow diagram in figure \ref{fig:flujo1} includes three distinct pathways marked with arrows for ease of interpretation. The grey arrow indicates that six students in the experimental condition formed initial groups of three members in the first semester and continued with the same members in the second semester. In the case of the control condition students formed initial groups of three in the first semester. Of these, eight students (indicated by the blue arrow) ended up in completely new groups in the second semester, while four (indicated by the green arrow) remained with two members of the original group. It's important to note that Figure \ref{fig:flujo1} does not indicate whether the new groups formed in the second semester were the same size as the original groups from the first semester. \\


After examining how students moved between groups, we used multiple regression models to explore the impact of the treatment on the number of team members a student chose to continue working with. The models in Table \ref{tab:r1} reveal that the treatment is a strong positive predictor (p $<$ 0.001) in all three models. Grade level also shows a positive and significant impact (p $<$ 0.01). Mean grades are significant predictors only in model 9. These findings suggest that students in the experimental group are more likely to keep their original team members than those in the control group. Additionally, 10th-grade students are also more inclined to maintain their initial group composition.

\begin{table}[!htbp] \centering 
\footnotesize	
  \caption{Regression models: Predicting the number of students with whom an individual continues to work when given the option to change groups.} 
  \label{tab:r1} 
\begin{tabular}{lccc} 
\cline{1-4}\\
 & \multicolumn{3}{c}{\textit{Dependent variable:}} \\ 
[0.5ex]\cline{2-4}
\\ & \multicolumn{3}{c}{n° students with whom a person continues to work.} \\ 
[0.5ex]\cline{2-4}
	\\&	 Model (9)	&	 Model (10)	&	 Model (11)	\\
 \hline \\ 
Exp. (Afinity-based groups)	&	1.69$^{***}$	&	1.67$^{***}$	&	1.22$^{***}$	\\
  	&	(0.16)	&	(0.17)	&	(0.34)	\\
 Group Size (3)     	&	-0.35	&	-0.34	&	-0.35	\\
                    	&	(0.23)	&	(0.24)	&	(0.23)	\\
 Gender (Female)    	&	-0.19	&	-0.21	&	-0.22	\\
                    	&	(0.15)	&	(0.16)	&	(0.16)	\\
  Grade Level (10th)	&	0.58$^{**}$	&	0.59$^{**}$	&	0.61$^{**}$	\\
 	&	(0.20)	&	(0.20)	&	(0.20)	\\
  Average grades    	&	0.02$^{*}$	&	0.02$^{\dagger}$	&	0.02$^{\dagger}$	\\
	&	(0.01)	&	(0.01)	&	(0.01)	\\
Friendship ($t_0$)	&		&	0.02	&	-0.04	\\
(in-degree)	&		&	(0.03)	&	(0.05)	\\
Exp. $*$ Friendship ($t_0$)	&		&		&	0.09	\\
(in-degree)	&		&		&	(0.06)	\\
Intercept	&	-1.38$^{\dagger}$	&	-1.35$^{\dagger}$	&	-1.04	\\
	&	(0.78)	&	(0.78)	&	(0.80)	\\
\hline \\ 											
Num.Obs.	&	90	&	90	&	90	\\
R2	&	0.63	&	0.63	&	0.64	\\
R2 Adj.	&	0.61	&	0.60	&	0.61	\\
AIC	&	205.2	&	206.8	&	206.4	\\
BIC	&	222.7	&	226.8	&	228.0	\\
Log.Lik.	&	-95.59	&	-95.39	&	-94.19	\\
F	&	28.32	&	23.48	&	20.73	\\
RMSE	&	0.72	&	0.73	&	0.72	\\

\\
\hline 

\textit{Note:}  & \multicolumn{3}{r}{$^{\dagger}$ p $<$ 0.1, $^{*}$ p $<$ 0.05, $^{**}$ p $<$ 0.01, $^{***}$p $<$ 0.001} 
\end{tabular} 
\end{table} 

\newpage

\section{Discussion}

The aim of the present study was to address two research questions. Firstly, we sought to investigate whether there were differences in team friendship, team cooperation, and the recognition of academic prestige between students in randomly assigned groups and those formed by self-selection after a semester of group work. Our second goal was to test whether these group configurations yield differences in the number of team members with whom students continue working after being given the chance to form new groups, after a semester of activities cooperating with their initial groups. To deepen in the comprehension of the observed differences between the experimental and control conditions, as well as across grade levels, we first examine the variations between the experimental and control groups, followed by an analysis focused on students' grade levels.

\subsection{Random and affinity-based groups}

Based on the comparative analysis of social network variables, we found, unsurprisingly that self-selected groups displayed a higher number of incoming friendship ties at the end of the semester compared to randomly assigned groups. Interestingly, groups assigned randomly achieved higher levels of cooperative ties during the same time period. Further, the number of friends before the intervention, both within the groups and across the classroom, were associated with team friendship following a semester of group work. Notably, initial team friendship did not relate to team cooperation, but it did to academic prestige within the group. Moreover, in relation to students' grade levels, older participants in both experimental and control conditions demonstrated elevated levels of team cooperation and friendship, presumably attributable to an additional year of shared experience, higher social awareness and/or maturity. Lastly, academic prestige was associated with average grades and gender, being males the ones who tend to enjoy such recognition. \\

Our results indicate that the number of team members with whom a student continues working from semester 1 to semester 2 was associated with the experimental condition (i.e., self-selected groups), and the grade level. Students initially randomly assigned to groups, who display lower levels of team friendship, but higher levels of team cooperation compared to those under the experimental condition, end up changing group partners more frequently when given the opportunity after semester 1. Interestingly, students in the control condition changed their team members even after achieving significantly higher cooperative group cohesion, yet less friendship cohesion than in the experimental cofigurations. The rate of membership change between control and experimental classrooms seems to add a contradicting extra layer of evidence to prior research. That is, previous studies had suggested that self-selected groups report higher levels of commitment, trust, satisfaction, and communication \cite{myersStudentsPerceptionsClassroom2012,chapmanCanWePick2006}. However, as we found, randomly assigned groups achieve higher team cooperative cohesion than those based on affinity, despite experiencing a higher rate of group change when given the opportunity. Conversely, when students have the autonomy to self-select their groups, they tend to form teams based on existing social connections, shared interests, or perceived compatibility, which can contribute to a more positive group dynamic \cite{myersStudentsPerceptionsClassroom2012}. According to our findings, self-selected groups demonstrated elevated levels of team friendship; however, this did not correspond to increased team cooperation, necessitating further investigation into the dynamics of cooperative and friendship relationships in academic settings.\\

Previous studies have found that students in randomly assigned groups may experience less social cohesion and emotional support than those in self-selected groups, probably leading to weaker group dynamics and a lower sense of belonging \cite{baumeisterNeedBelongDesire1995}. When students feel a strong sense of belonging and trust within their groups, they are more satisfied and motivated \cite{ennenImportanceTrustSatisfaction2015}, supporting the importance of a sense of belonging and engagement \cite{gillen-oneelSenseBelongingStudent2021}. Consequently, the absence of such feelings in randomly assigned groups could have reduced students' motivation to continue working together, despite their high cooperative cohesion. Thus, after one semester of working together, students under the control condition could have experienced the need to further explore alternative cooperative relationships based on affinity, for more effective processes towards group communication and coordination \cite{myersStudentsPerceptionsClassroom2012,chapmanCanWePick2006}. The fact that control groups developed higher team cooperative cohesion and yet, experienced the need to alternate group members might signal ineffective team dynamics for building a sense of belonging and trust, due to the lack of friendship. Furthermore, we have no information on whether this group change yields a positive/negative cooperative experience during semester 2 and, therefore, future research could address whether students' decisions for changing groups enable them an effective social and academic experience.\\

Another potential reason for our findings relates to the distribution of group roles and responsibilities. Research has indicated that when students are allowed to self-select their groups, they may be more likely to allocate roles based on individual strengths and preferences, leading to a more balanced and productive group dynamic \cite{myersStudentsPerceptionsClassroom2012}. On the contrary, randomly assigned groups might face group competition \cite{puurtinenJointEmergenceGroup2015}, potentially resulting in suboptimal teamwork and a lower likelihood of students wanting to continue working together, and overshadowing the potential benefits of their team cohesion. Furthermore, the differences in team cooperative cohesion might be the consequence of working alongside acquainted peers rather than friends, which could have fostered communication for addressing the various challenges associated with the assignments during the semester: assigning roles, division of tasks, coordinating efforts, decision making, among many others. The open nature of the activities might have amplified the need to learn the skills and weaknesses of the teammates, aiming for an effective coordination of labor \cite{davies2009, Steiner1966, pulgar_etal2020}. Such access to skill-related knowledge is complex among randomly assigned groups given the lack of existing relationship within the group, which could have encouraged more frequent communication among those involved and thus, higher team cooperative cohesion. Nonetheless, we could also argue that self-selected groups enjoy a level of social cohesion rooted in friendship, which might have hindered their cooperative efforts during the semester. Such a claim is consistent with the unintended consequences of high team cohesion, and associated with an overestimation of the team's capacities accompanied by an underestimation of the task's complexity \cite{parkEffectCohesionCurvilinear2017}.\\

While achievement metrics briefly influence the willingness to continue working with existing teammates, this effect becomes marginal after adding friendship variables. Students, particularly those in randomly assigned groups, overlook the learning gains afforded by their working units and make their assembling decisions based on relational information rather than on performance. This interpretation aligns with the dimensions of group effectiveness grounded on the premise that effective team processes promote psychological safety, thereby enhancing the desire to remain in the group \cite{hackman_2010}. That is, the peer selection process might have been conducted to find a safe place for communication and work, presumably a set of conditions easier to achieve among friends, as suggested above. Additionally, prior research on the longitudinal effects of collaboration on academic achievement in high school classrooms shows that working with friends becomes a more efficient interactive strategy over time compared to working with acquaintances, while these working ties remained stagnant in their association with grades as time progresses \cite{pulgar_etal_2022}. Consequently, individuals in the control condition might have experienced certain inadequacies of their cooperative dynamics in association with their performance, thus encouraging a group change with more familiar peers. Finally, it is important to note that the group selection strategy could respond to a level of affinity and/or competence not captured by the social network variables measured in this study.

\subsection{10th and 9th grade classrooms}

Tenth-grade participants exhibited a stronger inclination to continue collaborating with their first-semester teammates compared to those in the 9th grade. Individuals in 10th grade are older and presumably enjoy highly shared experiences compared to those in lower grades, and therefore, maturity and a more comprehensive awareness of their social relationships. This enhanced awareness likely fosters clearer social boundaries, enabling them to more effectively select classroom partners for an optimal cooperative experience. Accordingly, those who develop a more precise understanding of the surrounding social ties were better equipped to enact on social strategies for various purposes, like adjusting their behavior to the social norms of a particular group \cite{dunbarCognitiveConstraintsStructure2008}, or confronting others \cite{burtStructuralHolesSocial1995}. We posit that this enhanced exposure and experience in the classroom social system affords 10th graders knowledge and understanding of how others behave and respond in various scenarios, and consequently, could have led them to fortify their working relationships.\\

In a complementary interpretation, compared to 9 graders, 10th-grade students could be better equipped to face cooperative assignments that demand decision-making and group coordination due to the extra year of formal education where presumably, they gained experience and enhance their group-level competencies. Consequently, 10th-grade students would arguably be better prepared to work with acquainted peers, which would limit the tendency to change group members when given the opportunity. This interpretation aligns with research in psychology and education which points out that individuals prefer working with others who display cooperative behaviors \cite{johnsonActiveLearningCooperation2008,veselyProenvironmentalBehaviorSignal2020}. Conversely to younger students, who seem more prone to explore new cooperative ties for their group selection after one semester, 10th graders may have added higher value to their working experience during the semester for their grouping decision in semester 2. Here, it is worth noting that 10th grade students display statistically higher levels of team cooperation, compared to their younger counterparts. This maturity-driven interpretation could be worth exploring in future research. Finally, the above interpretations are consistent with previous research, which has demonstrated that cooperative groups tend to have more stability and are more likely to persist in working together over time \cite{gilliesCooperativeLearningReview2016,cavalierEffectsCooperativeLearning1995,johnsonEducationalPsychologySuccess2009}. Theories such as social interdependence theory \cite{johnsonSocialInterdependenceTheory2002} posit that positive interdependence (such as cooperative behavior) fosters commitment to the group and its tasks, leading to increased persistence. In a cooperative learning environment, students work together to achieve common goals, and this can foster a sense of belonging, increase motivation, and creates a collaborative culture \cite{johnsonStateCooperativeLearning2007a,johnsonLearningTogetherAlone2002}. These factors can subsequently increase the desire to remain in the group, as the members feel connected to each other and are committed to achieving their collective goals \cite{webbShortCircuitsSuperconductors2002}.

\subsection{Implication and limitations}

Given these findings, educators and curriculum designers should be mindful of the potential implications of cooperative teaching strategies and their effect on the learning environment. It becomes particularly important to promote, from early stages of school experience strategies for group cooperation, such as establishing clear group roles, responsibilities, encouraging open communication, and fostering a supportive learning environment. These guidelines may enhance group cohesion and students' desire to continue working together \cite{cohenWorkingEquityHeterogeneous1997, slavinResearchCooperativeLearning1996}, a sign of group effectiveness \cite{hackman_2010}. This can be achieved by setting clear boundaries for collaboration, providing structured opportunities for inter-group interactions, and monitoring the quality and quantity of external connections \cite{gersickTIMETRANSITIONWORK1988, tuckmanDevelopmentalSequenceSmall1965}. Nevertheless, further research is needed to explore the specific factors that drive these relationships and to identify optimal levels of internal and external cooperation that maximize learning outcomes and group satisfaction. Understanding the nuances of these dynamics will enable educators to create more effective learning environments that foster meaningful collaboration and to promote student success in both group and individual contexts. This could involve investigating the role of individual differences, such as personality traits or learning preferences, in shaping cooperation patterns and their impact on group dynamics \cite{felderLearningTeachingStyles2002a,costaFivefactorTheoryPersonality1999}, as well as examining the influence of contextual factors, such as group size or task complexity, on the effectiveness of cooperation strategies \cite{hackmanDesignWorkTeams1987, steinerGroupProcessProductivity1972}.

\section{Conclusions}

In summary, this study fills a gap in our understanding of how different methods of group formation affect collaboration, a key platform for developing 21st-century skills. Our main findings indicate that groups formed based on affinity (self-selection) were more likely to maintain their original composition over time compared to randomly assigned groups. Additionally, 10th-grade students were more inclined to keep their original group members. These affinity-based groups also showed higher levels of friendship but lower levels of cooperation compared to the control groups.\\

These results underscore the importance of creating educational environments that foster both cooperation and positive group dynamics to improve learning outcomes and group stability. However, the study has its limitations, including its focus on physics classes in a single school, the use of self-reported measures, and not considering other potentially influential factors like motivation and teamwork disposition. \\

Given these limitations, future research should aim to validate these findings in diverse educational settings and consider other variables that could influence group dynamics. Such research will contribute to the development of more effective teaching practices that not only enhance academic knowledge but also foster essential 21st-century skills.\\

\section*{Acknowledgement}

This study was suported by project FONDECYT 11220385, sponsored by Agencia Nacional de Investigación y Desarrollo (ANID), Chile. Cristian Candia was supported by FONDECYT Iniciaci\'on 11200986, sponsored by Agencia Nacional de Investigación y Desarrollo (ANID), Chile.

\bibliographystyle{apacite}
\bibliography{biblio}

\begin{thebibliography}{}

\bibitem [\protect \citeauthoryear {%
Bacon%
, Stewart%
\BCBL {}\ \BBA {} Silver%
}{%
Bacon%
\ \protect \BOthers {.}}{%
{\protect \APACyear {1999}}%
}]{%
baconLessonsBestWorst1999}
\APACinsertmetastar {%
baconLessonsBestWorst1999}%
\begin{APACrefauthors}%
Bacon, D\BPBI R.%
, Stewart, K\BPBI A.%
\BCBL {}\ \BBA {} Silver, W\BPBI S.%
\end{APACrefauthors}%
\unskip\
\newblock
\APACrefYearMonthDay{1999}{{\APACmonth{10}}}{}.
\newblock
{\BBOQ}\APACrefatitle {Lessons from the {{Best}} and {{Worst Student Team
  Experiences}}: {{How}} a {{Teacher}} Can Make the {{Difference}}} {Lessons
  from the {{Best}} and {{Worst Student Team Experiences}}: {{How}} a
  {{Teacher}} can make the {{Difference}}}.{\BBCQ}
\newblock
\APACjournalVolNumPages{Journal of Management Education}{23}{5}{467--488}.
\newblock
\begin{APACrefDOI} \doi{10.1177/105256299902300503} \end{APACrefDOI}
\PrintBackRefs{\CurrentBib}

\bibitem [\protect \citeauthoryear {%
Barron%
}{%
Barron%
}{%
{\protect \APACyear {2003}}%
}]{%
barronWhenSmartGroups2003}
\APACinsertmetastar {%
barronWhenSmartGroups2003}%
\begin{APACrefauthors}%
Barron, B.%
\end{APACrefauthors}%
\unskip\
\newblock
\APACrefYearMonthDay{2003}{{\APACmonth{07}}}{}.
\newblock
{\BBOQ}\APACrefatitle {When {{Smart Groups Fail}}} {When {{Smart Groups
  Fail}}}.{\BBCQ}
\newblock
\APACjournalVolNumPages{Journal of the Learning Sciences}{12}{3}{307--359}.
\newblock
\begin{APACrefDOI} \doi{10.1207/S15327809JLS1203_1} \end{APACrefDOI}
\PrintBackRefs{\CurrentBib}

\bibitem [\protect \citeauthoryear {%
Bastian%
, Heymann%
\BCBL {}\ \BBA {} Jacomy%
}{%
Bastian%
\ \protect \BOthers {.}}{%
{\protect \APACyear {2009}}%
}]{%
ICWSM09154}
\APACinsertmetastar {%
ICWSM09154}%
\begin{APACrefauthors}%
Bastian, M.%
, Heymann, S.%
\BCBL {}\ \BBA {} Jacomy, M.%
\end{APACrefauthors}%
\unskip\
\newblock
\APACrefYearMonthDay{2009}{}{}.
\newblock
\APACrefbtitle {Gephi: An Open Source Software for Exploring and Manipulating
  Networks.} {Gephi: An open source software for exploring and manipulating
  networks.}
\newblock
\begin{APACrefURL}
  \url{http://www.aaai.org/ocs/index.php/ICWSM/09/paper/view/154}
  \end{APACrefURL}
\PrintBackRefs{\CurrentBib}

\bibitem [\protect \citeauthoryear {%
Battistich%
, Solomon%
\BCBL {}\ \BBA {} Delucchi%
}{%
Battistich%
\ \protect \BOthers {.}}{%
{\protect \APACyear {1993}}%
}]{%
battistichInteractionProcessesStudent1993}
\APACinsertmetastar {%
battistichInteractionProcessesStudent1993}%
\begin{APACrefauthors}%
Battistich, V.%
, Solomon, D.%
\BCBL {}\ \BBA {} Delucchi, K.%
\end{APACrefauthors}%
\unskip\
\newblock
\APACrefYearMonthDay{1993}{{\APACmonth{09}}}{}.
\newblock
{\BBOQ}\APACrefatitle {Interaction {{Processes}} and {{Student Outcomes}} in
  {{Cooperative Learning Groups}}} {Interaction {{Processes}} and {{Student
  Outcomes}} in {{Cooperative Learning Groups}}}.{\BBCQ}
\newblock
\APACjournalVolNumPages{The Elementary School Journal}{94}{1}{19--32}.
\newblock
\begin{APACrefDOI} \doi{10.1086/461748} \end{APACrefDOI}
\PrintBackRefs{\CurrentBib}

\bibitem [\protect \citeauthoryear {%
Baumeister%
\ \BBA {} Leary%
}{%
Baumeister%
\ \BBA {} Leary%
}{%
{\protect \APACyear {1995}}%
}]{%
baumeisterNeedBelongDesire1995}
\APACinsertmetastar {%
baumeisterNeedBelongDesire1995}%
\begin{APACrefauthors}%
Baumeister, R\BPBI F.%
\BCBT {}\ \BBA {} Leary, M\BPBI R.%
\end{APACrefauthors}%
\unskip\
\newblock
\APACrefYearMonthDay{1995}{}{}.
\newblock
{\BBOQ}\APACrefatitle {The Need to Belong: {{Desire}} for Interpersonal
  Attachments as a Fundamental Human Motivation.} {The need to belong:
  {{Desire}} for interpersonal attachments as a fundamental human
  motivation.}{\BBCQ}
\newblock
\APACjournalVolNumPages{Psychological Bulletin}{117}{3}{497--529}.
\newblock
\begin{APACrefDOI} \doi{10.1037/0033-2909.117.3.497} \end{APACrefDOI}
\PrintBackRefs{\CurrentBib}

\bibitem [\protect \citeauthoryear {%
Blowers%
}{%
Blowers%
}{%
{\protect \APACyear {2003}}%
}]{%
blowersUsingStudentSkill2003}
\APACinsertmetastar {%
blowersUsingStudentSkill2003}%
\begin{APACrefauthors}%
Blowers, P.%
\end{APACrefauthors}%
\unskip\
\newblock
\APACrefYearMonthDay{2003}{{\APACmonth{08}}}{}.
\newblock
{\BBOQ}\APACrefatitle {Using {{Student Skill Self-Assessments}} to {{Get
  Balanced Groups}} for {{Group Projects}}} {Using {{Student Skill
  Self-Assessments}} to {{Get Balanced Groups}} for {{Group Projects}}}.{\BBCQ}
\newblock
\APACjournalVolNumPages{College Teaching}{51}{3}{106--110}.
\newblock
\begin{APACrefDOI} \doi{10.1080/87567550309596422} \end{APACrefDOI}
\PrintBackRefs{\CurrentBib}

\bibitem [\protect \citeauthoryear {%
Borgatti%
, Mehra%
, Brass%
\BCBL {}\ \BBA {} Labianca%
}{%
Borgatti%
\ \protect \BOthers {.}}{%
{\protect \APACyear {2009}}%
}]{%
borgattiNetworkAnalysisSocial2009a}
\APACinsertmetastar {%
borgattiNetworkAnalysisSocial2009a}%
\begin{APACrefauthors}%
Borgatti, S\BPBI P.%
, Mehra, A.%
, Brass, D\BPBI J.%
\BCBL {}\ \BBA {} Labianca, G.%
\end{APACrefauthors}%
\unskip\
\newblock
\APACrefYearMonthDay{2009}{{\APACmonth{02}}}{}.
\newblock
{\BBOQ}\APACrefatitle {Network {{Analysis}} in the {{Social Sciences}}}
  {Network {{Analysis}} in the {{Social Sciences}}}.{\BBCQ}
\newblock
\APACjournalVolNumPages{Science}{323}{5916}{892--895}.
\newblock
\begin{APACrefDOI} \doi{10.1126/science.1165821} \end{APACrefDOI}
\PrintBackRefs{\CurrentBib}

\bibitem [\protect \citeauthoryear {%
Brown%
\ \BBA {} Paulus%
}{%
Brown%
\ \BBA {} Paulus%
}{%
{\protect \APACyear {2002}}%
}]{%
brownMakingGroupBrainstorming2002}
\APACinsertmetastar {%
brownMakingGroupBrainstorming2002}%
\begin{APACrefauthors}%
Brown, V\BPBI R.%
\BCBT {}\ \BBA {} Paulus, P\BPBI B.%
\end{APACrefauthors}%
\unskip\
\newblock
\APACrefYearMonthDay{2002}{{\APACmonth{12}}}{}.
\newblock
{\BBOQ}\APACrefatitle {Making {{Group Brainstorming More Effective}}:
  {{Recommendations From}} an {{Associative Memory Perspective}}} {Making
  {{Group Brainstorming More Effective}}: {{Recommendations From}} an
  {{Associative Memory Perspective}}}.{\BBCQ}
\newblock
\APACjournalVolNumPages{Current Directions in Psychological
  Science}{11}{6}{208--212}.
\newblock
\begin{APACrefDOI} \doi{10.1111/1467-8721.00202} \end{APACrefDOI}
\PrintBackRefs{\CurrentBib}

\bibitem [\protect \citeauthoryear {%
Burt%
}{%
Burt%
}{%
{\protect \APACyear {1995}}%
}]{%
burtStructuralHolesSocial1995}
\APACinsertmetastar {%
burtStructuralHolesSocial1995}%
\begin{APACrefauthors}%
Burt, R\BPBI S.%
\end{APACrefauthors}%
\unskip\
\newblock
\APACrefYear{1995}.
\newblock
\APACrefbtitle {Structural holes: the social structure of competition}
  {Structural holes: the social structure of competition}\ (\PrintOrdinal{1.
  Harvard Univ. Press paperback ed}\ \BEd).
\newblock
\APACaddressPublisher{Cambridge, Mass.}{Harvard University Press}.
\PrintBackRefs{\CurrentBib}

\bibitem [\protect \citeauthoryear {%
Candia%
, Oyarzún%
\BCBL {}\ \protect \BOthers {.}}{%
Candia%
, Oyarzún%
\BCBL {}\ \protect \BOthers {.}}{%
{\protect \APACyear {2022}}%
}]{%
candia2022reciprocity}
\APACinsertmetastar {%
candia2022reciprocity}%
\begin{APACrefauthors}%
Candia, C.%
, Oyarzún, M.%
, Landaeta, V.%
, Yaikin, T.%
, Monge, C.%
, Hidalgo, C.%
\BCBL {}\ \BBA {} Rodriguez-Sickert, C.%
\end{APACrefauthors}%
\unskip\
\newblock
\APACrefYearMonthDay{2022}{}{}.
\newblock
{\BBOQ}\APACrefatitle {Reciprocity heightens academic performance in elementary
  school students} {Reciprocity heightens academic performance in elementary
  school students}.{\BBCQ}
\newblock
\APACjournalVolNumPages{Heliyon}{8}{12}{e11916}.
\newblock
\begin{APACrefURL}
  \url{https://www.sciencedirect.com/science/article/pii/S2405844022032042}
  \end{APACrefURL}
\newblock
\begin{APACrefDOI} \doi{https://doi.org/10.1016/j.heliyon.2022.e11916}
  \end{APACrefDOI}
\PrintBackRefs{\CurrentBib}

\bibitem [\protect \citeauthoryear {%
Candia%
, Pulgar%
\BCBL {}\ \BBA {} Pinheiro%
}{%
Candia%
, Pulgar%
\BCBL {}\ \BBA {} Pinheiro%
}{%
{\protect \APACyear {2022}}%
}]{%
candia2022interconnectedness}
\APACinsertmetastar {%
candia2022interconnectedness}%
\begin{APACrefauthors}%
Candia, C.%
, Pulgar, J.%
\BCBL {}\ \BBA {} Pinheiro, F.%
\end{APACrefauthors}%
\unskip\
\newblock
\APACrefYearMonthDay{2022}{}{}.
\newblock
{\BBOQ}\APACrefatitle {Interconnectedness in Education Systems}
  {Interconnectedness in education systems}.{\BBCQ}
\newblock
\APACjournalVolNumPages{arXiv preprint arXiv:2203.05624}{}{}{}.
\PrintBackRefs{\CurrentBib}

\bibitem [\protect \citeauthoryear {%
Carmeli%
, Gelbard%
\BCBL {}\ \BBA {} Goldriech%
}{%
Carmeli%
\ \protect \BOthers {.}}{%
{\protect \APACyear {2011}}%
}]{%
carmeliLinkingPerceivedExternal2011a}
\APACinsertmetastar {%
carmeliLinkingPerceivedExternal2011a}%
\begin{APACrefauthors}%
Carmeli, A.%
, Gelbard, R.%
\BCBL {}\ \BBA {} Goldriech, R.%
\end{APACrefauthors}%
\unskip\
\newblock
\APACrefYearMonthDay{2011}{{\APACmonth{07}}}{}.
\newblock
{\BBOQ}\APACrefatitle {Linking Perceived External Prestige and Collective
  Identification to Collaborative Behaviors in {{R}}\&{{D}} Teams} {Linking
  perceived external prestige and collective identification to collaborative
  behaviors in {{R}}\&{{D}} teams}.{\BBCQ}
\newblock
\APACjournalVolNumPages{Expert Systems with Applications}{38}{7}{8199--8207}.
\newblock
\begin{APACrefDOI} \doi{10.1016/j.eswa.2010.12.166} \end{APACrefDOI}
\PrintBackRefs{\CurrentBib}

\bibitem [\protect \citeauthoryear {%
Cavalier%
, Klein%
\BCBL {}\ \BBA {} Cavalier%
}{%
Cavalier%
\ \protect \BOthers {.}}{%
{\protect \APACyear {1995}}%
}]{%
cavalierEffectsCooperativeLearning1995}
\APACinsertmetastar {%
cavalierEffectsCooperativeLearning1995}%
\begin{APACrefauthors}%
Cavalier, J\BPBI C.%
, Klein, J\BPBI D.%
\BCBL {}\ \BBA {} Cavalier, F\BPBI J.%
\end{APACrefauthors}%
\unskip\
\newblock
\APACrefYearMonthDay{1995}{{\APACmonth{09}}}{}.
\newblock
{\BBOQ}\APACrefatitle {Effects of Cooperative Learning on Performance,
  Attitude, and Group Behaviors in a Technical Team Environment} {Effects of
  cooperative learning on performance, attitude, and group behaviors in a
  technical team environment}.{\BBCQ}
\newblock
\APACjournalVolNumPages{Educational Technology Research and
  Development}{43}{3}{61--72}.
\newblock
\begin{APACrefDOI} \doi{10.1007/BF02300456} \end{APACrefDOI}
\PrintBackRefs{\CurrentBib}

\bibitem [\protect \citeauthoryear {%
Cestone%
, Levine%
\BCBL {}\ \BBA {} Lane%
}{%
Cestone%
\ \protect \BOthers {.}}{%
{\protect \APACyear {2008}}%
}]{%
cestonePeerAssessmentEvaluation2008}
\APACinsertmetastar {%
cestonePeerAssessmentEvaluation2008}%
\begin{APACrefauthors}%
Cestone, C\BPBI M.%
, Levine, R\BPBI E.%
\BCBL {}\ \BBA {} Lane, D\BPBI R.%
\end{APACrefauthors}%
\unskip\
\newblock
\APACrefYearMonthDay{2008}{{\APACmonth{09}}}{}.
\newblock
{\BBOQ}\APACrefatitle {Peer Assessment and Evaluation in Team-Based Learning}
  {Peer assessment and evaluation in team-based learning}.{\BBCQ}
\newblock
\APACjournalVolNumPages{New Directions for Teaching and
  Learning}{2008}{116}{69--78}.
\newblock
\begin{APACrefDOI} \doi{10.1002/tl.334} \end{APACrefDOI}
\PrintBackRefs{\CurrentBib}

\bibitem [\protect \citeauthoryear {%
Chapman%
, Meuter%
, Toy%
\BCBL {}\ \BBA {} Wright%
}{%
Chapman%
\ \protect \BOthers {.}}{%
{\protect \APACyear {2006}}%
}]{%
chapmanCanWePick2006}
\APACinsertmetastar {%
chapmanCanWePick2006}%
\begin{APACrefauthors}%
Chapman, K\BPBI J.%
, Meuter, M.%
, Toy, D.%
\BCBL {}\ \BBA {} Wright, L.%
\end{APACrefauthors}%
\unskip\
\newblock
\APACrefYearMonthDay{2006}{{\APACmonth{08}}}{}.
\newblock
{\BBOQ}\APACrefatitle {Can't {{We Pick}} Our {{Own Groups}}? {{The Influence}}
  of {{Group Selection Method}} on {{Group Dynamics}} and {{Outcomes}}} {Can't
  {{We Pick}} our {{Own Groups}}? {{The Influence}} of {{Group Selection
  Method}} on {{Group Dynamics}} and {{Outcomes}}}.{\BBCQ}
\newblock
\APACjournalVolNumPages{Journal of Management Education}{30}{4}{557--569}.
\newblock
\begin{APACrefDOI} \doi{10.1177/1052562905284872} \end{APACrefDOI}
\PrintBackRefs{\CurrentBib}

\bibitem [\protect \citeauthoryear {%
Cho%
, Gay%
, Davidson%
\BCBL {}\ \BBA {} Ingraffea%
}{%
Cho%
\ \protect \BOthers {.}}{%
{\protect \APACyear {2007}}%
}]{%
choSocialNetworksCommunication2007b}
\APACinsertmetastar {%
choSocialNetworksCommunication2007b}%
\begin{APACrefauthors}%
Cho, H.%
, Gay, G.%
, Davidson, B.%
\BCBL {}\ \BBA {} Ingraffea, A.%
\end{APACrefauthors}%
\unskip\
\newblock
\APACrefYearMonthDay{2007}{{\APACmonth{09}}}{}.
\newblock
{\BBOQ}\APACrefatitle {Social Networks, Communication Styles, and Learning
  Performance in a {{CSCL}} Community} {Social networks, communication styles,
  and learning performance in a {{CSCL}} community}.{\BBCQ}
\newblock
\APACjournalVolNumPages{Computers \& Education}{49}{2}{309--329}.
\newblock
\begin{APACrefDOI} \doi{10.1016/j.compedu.2005.07.003} \end{APACrefDOI}
\PrintBackRefs{\CurrentBib}

\bibitem [\protect \citeauthoryear {%
Cohen%
}{%
Cohen%
}{%
{\protect \APACyear {1994}}%
}]{%
cohenRestructuringClassroomConditions1994}
\APACinsertmetastar {%
cohenRestructuringClassroomConditions1994}%
\begin{APACrefauthors}%
Cohen, E\BPBI G.%
\end{APACrefauthors}%
\unskip\
\newblock
\APACrefYearMonthDay{1994}{{\APACmonth{03}}}{}.
\newblock
{\BBOQ}\APACrefatitle {Restructuring the {{Classroom}}: {{Conditions}} for
  {{Productive Small Groups}}} {Restructuring the {{Classroom}}: {{Conditions}}
  for {{Productive Small Groups}}}.{\BBCQ}
\newblock
\APACjournalVolNumPages{Review of Educational Research}{64}{1}{1--35}.
\newblock
\begin{APACrefDOI} \doi{10.3102/00346543064001001} \end{APACrefDOI}
\PrintBackRefs{\CurrentBib}

\bibitem [\protect \citeauthoryear {%
Cohen%
\ \BBA {} Lotan%
}{%
Cohen%
\ \BBA {} Lotan%
}{%
{\protect \APACyear {1997}}%
}]{%
cohenWorkingEquityHeterogeneous1997}
\APACinsertmetastar {%
cohenWorkingEquityHeterogeneous1997}%
\begin{APACrefauthors}%
Cohen, E\BPBI G.%
\BCBT {}\ \BBA {} Lotan, R\BPBI A.%
\end{APACrefauthors}%
\unskip\
\newblock
\APACrefYear{1997}.
\newblock
\APACrefbtitle {Working for Equity in Heterogeneous Classrooms:
  {{Sociological}} Theory in Practice} {Working for equity in heterogeneous
  classrooms: {{Sociological}} theory in practice}.
\newblock
\APACaddressPublisher{}{{Teachers College Press}}.
\PrintBackRefs{\CurrentBib}

\bibitem [\protect \citeauthoryear {%
Connerley%
\ \BBA {} Mael%
}{%
Connerley%
\ \BBA {} Mael%
}{%
{\protect \APACyear {2001}}%
}]{%
connerleyImportanceInvasivenessStudent2001}
\APACinsertmetastar {%
connerleyImportanceInvasivenessStudent2001}%
\begin{APACrefauthors}%
Connerley, M\BPBI L.%
\BCBT {}\ \BBA {} Mael, F\BPBI A.%
\end{APACrefauthors}%
\unskip\
\newblock
\APACrefYearMonthDay{2001}{{\APACmonth{10}}}{}.
\newblock
{\BBOQ}\APACrefatitle {The {{Importance}} and {{Invasiveness}} of {{Student
  Team Selection Criteria}}} {The {{Importance}} and {{Invasiveness}} of
  {{Student Team Selection Criteria}}}.{\BBCQ}
\newblock
\APACjournalVolNumPages{Journal of Management Education}{25}{5}{471--494}.
\newblock
\begin{APACrefDOI} \doi{10.1177/105256290102500502} \end{APACrefDOI}
\PrintBackRefs{\CurrentBib}

\bibitem [\protect \citeauthoryear {%
Costa%
\ \BBA {} McCrae%
}{%
Costa%
\ \BBA {} McCrae%
}{%
{\protect \APACyear {1999}}%
}]{%
costaFivefactorTheoryPersonality1999}
\APACinsertmetastar {%
costaFivefactorTheoryPersonality1999}%
\begin{APACrefauthors}%
Costa, P\BPBI T.%
\BCBT {}\ \BBA {} McCrae, R\BPBI R.%
\end{APACrefauthors}%
\unskip\
\newblock
\APACrefYearMonthDay{1999}{}{}.
\newblock
{\BBOQ}\APACrefatitle {A Five-Factor Theory of Personality} {A five-factor
  theory of personality}.{\BBCQ}
\newblock
\APACjournalVolNumPages{The five-factor model of personality: Theoretical
  perspectives}{2}{}{51--87}.
\PrintBackRefs{\CurrentBib}

\bibitem [\protect \citeauthoryear {%
Davies%
}{%
Davies%
}{%
{\protect \APACyear {2009}}%
}]{%
davies2009}
\APACinsertmetastar {%
davies2009}%
\begin{APACrefauthors}%
Davies, M\BPBI W.%
\end{APACrefauthors}%
\unskip\
\newblock
\APACrefYearMonthDay{2009}{}{}.
\newblock
{\BBOQ}\APACrefatitle {Group work as a form of assessment: common problems and
  recommended solutions} {Group work as a form of assessment: common problems
  and recommended solutions}.{\BBCQ}
\newblock
\APACjournalVolNumPages{Higher Education}{58}{}{563-584}.
\PrintBackRefs{\CurrentBib}

\bibitem [\protect \citeauthoryear {%
Deibel%
}{%
Deibel%
}{%
{\protect \APACyear {2005}}%
}]{%
deibelTeamFormationMethods2005}
\APACinsertmetastar {%
deibelTeamFormationMethods2005}%
\begin{APACrefauthors}%
Deibel, K.%
\end{APACrefauthors}%
\unskip\
\newblock
\APACrefYearMonthDay{2005}{{\APACmonth{06}}}{}.
\newblock
{\BBOQ}\APACrefatitle {Team Formation Methods for Increasing Interaction during
  In-Class Group Work} {Team formation methods for increasing interaction
  during in-class group work}.{\BBCQ}
\newblock
\BIn{} \APACrefbtitle {Proceedings of the 10th Annual {{SIGCSE}} Conference on
  {{Innovation}} and Technology in Computer Science Education} {Proceedings of
  the 10th annual {{SIGCSE}} conference on {{Innovation}} and technology in
  computer science education}\ (\BPGS\ 291--295).
\newblock
\APACaddressPublisher{{Caparica Portugal}}{{ACM}}.
\newblock
\begin{APACrefDOI} \doi{10.1145/1067445.1067525} \end{APACrefDOI}
\PrintBackRefs{\CurrentBib}

\bibitem [\protect \citeauthoryear {%
Dou%
\ \BBA {} Zwolak%
}{%
Dou%
\ \BBA {} Zwolak%
}{%
{\protect \APACyear {2019}}%
}]{%
douPractitionerGuideSocial2019}
\APACinsertmetastar {%
douPractitionerGuideSocial2019}%
\begin{APACrefauthors}%
Dou, R.%
\BCBT {}\ \BBA {} Zwolak, J\BPBI P.%
\end{APACrefauthors}%
\unskip\
\newblock
\APACrefYearMonthDay{2019}{{\APACmonth{07}}}{}.
\newblock
{\BBOQ}\APACrefatitle {Practitioner's Guide to Social Network Analysis:
  {{Examining}} Physics Anxiety in an Active-Learning Setting} {Practitioner's
  guide to social network analysis: {{Examining}} physics anxiety in an
  active-learning setting}.{\BBCQ}
\newblock
\APACjournalVolNumPages{Physical Review Physics Education
  Research}{15}{2}{020105}.
\newblock
\begin{APACrefDOI} \doi{10.1103/PhysRevPhysEducRes.15.020105} \end{APACrefDOI}
\PrintBackRefs{\CurrentBib}

\bibitem [\protect \citeauthoryear {%
Dunbar%
}{%
Dunbar%
}{%
{\protect \APACyear {2008}}%
}]{%
dunbarCognitiveConstraintsStructure2008}
\APACinsertmetastar {%
dunbarCognitiveConstraintsStructure2008}%
\begin{APACrefauthors}%
Dunbar, R\BPBI I\BPBI M.%
\end{APACrefauthors}%
\unskip\
\newblock
\APACrefYearMonthDay{2008}{}{}.
\newblock
{\BBOQ}\APACrefatitle {Cognitive constraints on the structure and dynamics of
  social networks.} {Cognitive constraints on the structure and dynamics of
  social networks.}{\BBCQ}
\newblock
\APACjournalVolNumPages{Group Dynamics: Theory, Research, and
  Practice}{12}{1}{7--16}.
\newblock
\begin{APACrefURL}
  [{2022-06-04}]\url{http://doi.apa.org/getdoi.cfm?doi=10.1037/1089-2699.12.1.7}
  \end{APACrefURL}
\newblock
\begin{APACrefDOI} \doi{10.1037/1089-2699.12.1.7} \end{APACrefDOI}
\PrintBackRefs{\CurrentBib}

\bibitem [\protect \citeauthoryear {%
Elias%
}{%
Elias%
}{%
{\protect \APACyear {1997}}%
}]{%
eliasPromotingSocialEmotional1997}
\APACinsertmetastar {%
eliasPromotingSocialEmotional1997}%
\begin{APACrefauthors}%
Elias, M\BPBI J.%
\end{APACrefauthors}%
\unskip\
\newblock
\APACrefYear{1997}.
\newblock
\APACrefbtitle {Promoting {{Social}} and {{Emotional Learning}}: {{Guidelines}}
  for {{Educators}}} {Promoting {{Social}} and {{Emotional Learning}}:
  {{Guidelines}} for {{Educators}}}.
\newblock
\APACaddressPublisher{}{{ASCD}}.
\PrintBackRefs{\CurrentBib}

\bibitem [\protect \citeauthoryear {%
Ennen%
, Stark%
\BCBL {}\ \BBA {} Lassiter%
}{%
Ennen%
\ \protect \BOthers {.}}{%
{\protect \APACyear {2015}}%
}]{%
ennenImportanceTrustSatisfaction2015}
\APACinsertmetastar {%
ennenImportanceTrustSatisfaction2015}%
\begin{APACrefauthors}%
Ennen, N\BPBI L.%
, Stark, E.%
\BCBL {}\ \BBA {} Lassiter, A.%
\end{APACrefauthors}%
\unskip\
\newblock
\APACrefYearMonthDay{2015}{{\APACmonth{09}}}{}.
\newblock
{\BBOQ}\APACrefatitle {The Importance of Trust for Satisfaction, Motivation,
  and Academic Performance in Student Learning Groups} {The importance of trust
  for satisfaction, motivation, and academic performance in student learning
  groups}.{\BBCQ}
\newblock
\APACjournalVolNumPages{Social Psychology of Education}{18}{3}{615--633}.
\newblock
\begin{APACrefDOI} \doi{10.1007/s11218-015-9306-x} \end{APACrefDOI}
\PrintBackRefs{\CurrentBib}

\bibitem [\protect \citeauthoryear {%
Felder%
}{%
Felder%
}{%
{\protect \APACyear {2002}}%
}]{%
felderLearningTeachingStyles2002a}
\APACinsertmetastar {%
felderLearningTeachingStyles2002a}%
\begin{APACrefauthors}%
Felder, R\BPBI M.%
\end{APACrefauthors}%
\unskip\
\newblock
\APACrefYearMonthDay{2002}{}{}.
\newblock
{\BBOQ}\APACrefatitle {Learning and Teaching Styles in Engineering Education}
  {Learning and teaching styles in engineering education}.{\BBCQ}
\newblock

\PrintBackRefs{\CurrentBib}

\bibitem [\protect \citeauthoryear {%
Festinger%
}{%
Festinger%
}{%
{\protect \APACyear {1950}}%
}]{%
festingerInformalSocialCommunication1950}
\APACinsertmetastar {%
festingerInformalSocialCommunication1950}%
\begin{APACrefauthors}%
Festinger, L.%
\end{APACrefauthors}%
\unskip\
\newblock
\APACrefYearMonthDay{1950}{{\APACmonth{09}}}{}.
\newblock
{\BBOQ}\APACrefatitle {Informal Social Communication.} {Informal social
  communication.}{\BBCQ}
\newblock
\APACjournalVolNumPages{Psychological Review}{57}{5}{271--282}.
\newblock
\begin{APACrefDOI} \doi{10.1037/h0056932} \end{APACrefDOI}
\PrintBackRefs{\CurrentBib}

\bibitem [\protect \citeauthoryear {%
Fiechtner%
\ \BBA {} Davis%
}{%
Fiechtner%
\ \BBA {} Davis%
}{%
{\protect \APACyear {2016}}%
}]{%
fiechtnerRepublicationWhyGroups2016a}
\APACinsertmetastar {%
fiechtnerRepublicationWhyGroups2016a}%
\begin{APACrefauthors}%
Fiechtner, S\BPBI B.%
\BCBT {}\ \BBA {} Davis, E\BPBI A.%
\end{APACrefauthors}%
\unskip\
\newblock
\APACrefYearMonthDay{2016}{{\APACmonth{02}}}{}.
\newblock
{\BBOQ}\APACrefatitle {Republication of ``{{Why}} Some Groups Fail: {{A}}
  Survey of Students' Experiences with Learning Groups''} {Republication of
  ``{{Why}} some groups fail: {{A}} survey of students' experiences with
  learning groups''}.{\BBCQ}
\newblock
\APACjournalVolNumPages{Journal of Management Education}{40}{1}{12--29}.
\newblock
\begin{APACrefDOI} \doi{10.1177/1052562915619639} \end{APACrefDOI}
\PrintBackRefs{\CurrentBib}

\bibitem [\protect \citeauthoryear {%
Ga{\v s}evi{\'c}%
, Zouaq%
\BCBL {}\ \BBA {} Janzen%
}{%
Ga{\v s}evi{\'c}%
\ \protect \BOthers {.}}{%
{\protect \APACyear {2013}}%
}]{%
gasevicChooseYourClassmates2013}
\APACinsertmetastar {%
gasevicChooseYourClassmates2013}%
\begin{APACrefauthors}%
Ga{\v s}evi{\'c}, D.%
, Zouaq, A.%
\BCBL {}\ \BBA {} Janzen, R.%
\end{APACrefauthors}%
\unskip\
\newblock
\APACrefYearMonthDay{2013}{{\APACmonth{10}}}{}.
\newblock
{\BBOQ}\APACrefatitle {``{{Choose Your Classmates}}, {{Your GPA Is}} at
  {{Stake}}!'': {{The Association}} of {{Cross-Class Social Ties}} and
  {{Academic Performance}}} {``{{Choose Your Classmates}}, {{Your GPA Is}} at
  {{Stake}}!'': {{The Association}} of {{Cross-Class Social Ties}} and
  {{Academic Performance}}}.{\BBCQ}
\newblock
\APACjournalVolNumPages{American Behavioral Scientist}{57}{10}{1460--1479}.
\newblock
\begin{APACrefDOI} \doi{10.1177/0002764213479362} \end{APACrefDOI}
\PrintBackRefs{\CurrentBib}

\bibitem [\protect \citeauthoryear {%
Gersick%
}{%
Gersick%
}{%
{\protect \APACyear {1988}}%
}]{%
gersickTIMETRANSITIONWORK1988}
\APACinsertmetastar {%
gersickTIMETRANSITIONWORK1988}%
\begin{APACrefauthors}%
Gersick, C\BPBI J\BPBI G.%
\end{APACrefauthors}%
\unskip\
\newblock
\APACrefYearMonthDay{1988}{{\APACmonth{03}}}{}.
\newblock
{\BBOQ}\APACrefatitle {{{TIME AN TRANSITION IN WORK TEAMS}}: {{TOWARD A NEW
  MODEL OF GROUP DEVELOPMENT}}.} {{{TIME AN TRANSITION IN WORK TEAMS}}:
  {{TOWARD A NEW MODEL OF GROUP DEVELOPMENT}}.}{\BBCQ}
\newblock
\APACjournalVolNumPages{Academy of Management Journal}{31}{1}{9--41}.
\newblock
\begin{APACrefDOI} \doi{10.2307/256496} \end{APACrefDOI}
\PrintBackRefs{\CurrentBib}

\bibitem [\protect \citeauthoryear {%
Getzels%
\ \BBA {} Thelen%
}{%
Getzels%
\ \BBA {} Thelen%
}{%
{\protect \APACyear {1960}}%
}]{%
getzelsClassroomGroupUnique1960}
\APACinsertmetastar {%
getzelsClassroomGroupUnique1960}%
\begin{APACrefauthors}%
Getzels, J\BPBI W.%
\BCBT {}\ \BBA {} Thelen, H\BPBI A.%
\end{APACrefauthors}%
\unskip\
\newblock
\APACrefYearMonthDay{1960}{{\APACmonth{08}}}{}.
\newblock
{\BBOQ}\APACrefatitle {The {{Classroom Group}} as a {{Unique Social System}}}
  {The {{Classroom Group}} as a {{Unique Social System}}}.{\BBCQ}
\newblock
\APACjournalVolNumPages{Teachers College Record}{61}{10}{53--82}.
\newblock
\begin{APACrefDOI} \doi{10.1177/016146816006101004} \end{APACrefDOI}
\PrintBackRefs{\CurrentBib}

\bibitem [\protect \citeauthoryear {%
{Gillen-O'Neel}%
}{%
{Gillen-O'Neel}%
}{%
{\protect \APACyear {2021}}%
}]{%
gillen-oneelSenseBelongingStudent2021}
\APACinsertmetastar {%
gillen-oneelSenseBelongingStudent2021}%
\begin{APACrefauthors}%
{Gillen-O'Neel}, C.%
\end{APACrefauthors}%
\unskip\
\newblock
\APACrefYearMonthDay{2021}{{\APACmonth{02}}}{}.
\newblock
{\BBOQ}\APACrefatitle {Sense of {{Belonging}} and {{Student Engagement}}: {{A
  Daily Study}} of {{First-}} and {{Continuing-Generation College Students}}}
  {Sense of {{Belonging}} and {{Student Engagement}}: {{A Daily Study}} of
  {{First-}} and {{Continuing-Generation College Students}}}.{\BBCQ}
\newblock
\APACjournalVolNumPages{Research in Higher Education}{62}{1}{45--71}.
\newblock
\begin{APACrefDOI} \doi{10.1007/s11162-019-09570-y} \end{APACrefDOI}
\PrintBackRefs{\CurrentBib}

\bibitem [\protect \citeauthoryear {%
Gillies%
}{%
Gillies%
}{%
{\protect \APACyear {2003}}%
}]{%
gilliesBehaviorsInteractionsPerceptions2003}
\APACinsertmetastar {%
gilliesBehaviorsInteractionsPerceptions2003}%
\begin{APACrefauthors}%
Gillies, R\BPBI M.%
\end{APACrefauthors}%
\unskip\
\newblock
\APACrefYearMonthDay{2003}{{\APACmonth{03}}}{}.
\newblock
{\BBOQ}\APACrefatitle {The Behaviors, Interactions, and Perceptions of Junior
  High School Students during Small-Group Learning.} {The behaviors,
  interactions, and perceptions of junior high school students during
  small-group learning.}{\BBCQ}
\newblock
\APACjournalVolNumPages{Journal of Educational Psychology}{95}{1}{137--147}.
\newblock
\begin{APACrefDOI} \doi{10.1037/0022-0663.95.1.137} \end{APACrefDOI}
\PrintBackRefs{\CurrentBib}

\bibitem [\protect \citeauthoryear {%
Gillies%
}{%
Gillies%
}{%
{\protect \APACyear {2004}}%
}]{%
gilliesEffectsCooperativeLearning2004}
\APACinsertmetastar {%
gilliesEffectsCooperativeLearning2004}%
\begin{APACrefauthors}%
Gillies, R\BPBI M.%
\end{APACrefauthors}%
\unskip\
\newblock
\APACrefYearMonthDay{2004}{{\APACmonth{04}}}{}.
\newblock
{\BBOQ}\APACrefatitle {The Effects of Cooperative Learning on Junior High
  School Students during Small Group Learning} {The effects of cooperative
  learning on junior high school students during small group learning}.{\BBCQ}
\newblock
\APACjournalVolNumPages{Learning and Instruction}{14}{2}{197--213}.
\newblock
\begin{APACrefDOI} \doi{10.1016/S0959-4752(03)00068-9} \end{APACrefDOI}
\PrintBackRefs{\CurrentBib}

\bibitem [\protect \citeauthoryear {%
Gillies%
}{%
Gillies%
}{%
{\protect \APACyear {2016}}%
}]{%
gilliesCooperativeLearningReview2016}
\APACinsertmetastar {%
gilliesCooperativeLearningReview2016}%
\begin{APACrefauthors}%
Gillies, R\BPBI M.%
\end{APACrefauthors}%
\unskip\
\newblock
\APACrefYearMonthDay{2016}{}{}.
\newblock
{\BBOQ}\APACrefatitle {Cooperative Learning: {{Review}} of Research and
  Practice} {Cooperative learning: {{Review}} of research and practice}.{\BBCQ}
\newblock
\APACjournalVolNumPages{Australian Journal of Teacher Education
  (Online)}{41}{3}{39--54}.
\PrintBackRefs{\CurrentBib}

\bibitem [\protect \citeauthoryear {%
Goswami%
\ \BBA {} Agrawal%
}{%
Goswami%
\ \BBA {} Agrawal%
}{%
{\protect \APACyear {2020}}%
}]{%
goswamiExplicatingInfluenceShared2020}
\APACinsertmetastar {%
goswamiExplicatingInfluenceShared2020}%
\begin{APACrefauthors}%
Goswami, A\BPBI K.%
\BCBT {}\ \BBA {} Agrawal, R\BPBI K.%
\end{APACrefauthors}%
\unskip\
\newblock
\APACrefYearMonthDay{2020}{{\APACmonth{03}}}{}.
\newblock
{\BBOQ}\APACrefatitle {Explicating the Influence of Shared Goals and Hope on
  Knowledge Sharing and Knowledge Creation in an Emerging Economic Context}
  {Explicating the influence of shared goals and hope on knowledge sharing and
  knowledge creation in an emerging economic context}.{\BBCQ}
\newblock
\APACjournalVolNumPages{Journal of Knowledge Management}{24}{2}{172--195}.
\newblock
\begin{APACrefDOI} \doi{10.1108/JKM-09-2018-0561} \end{APACrefDOI}
\PrintBackRefs{\CurrentBib}

\bibitem [\protect \citeauthoryear {%
Hackman%
}{%
Hackman%
}{%
{\protect \APACyear {1987}}%
}]{%
hackmanDesignWorkTeams1987}
\APACinsertmetastar {%
hackmanDesignWorkTeams1987}%
\begin{APACrefauthors}%
Hackman, J\BPBI R.%
\end{APACrefauthors}%
\unskip\
\newblock
\APACrefYearMonthDay{1987}{}{}.
\newblock
{\BBOQ}\APACrefatitle {The Design of Work Teams. {{InHandbook}} of
  {{Organizational Behavior}}, Ed. {{JW Lorsch}}} {The design of work teams.
  {{InHandbook}} of {{Organizational Behavior}}, ed. {{JW Lorsch}}}.{\BBCQ}
\newblock

\PrintBackRefs{\CurrentBib}

\bibitem [\protect \citeauthoryear {%
Hackman%
\ \BBA {} Katz%
}{%
Hackman%
\ \BBA {} Katz%
}{%
{\protect \APACyear {2010}}%
}]{%
hackman_2010}
\APACinsertmetastar {%
hackman_2010}%
\begin{APACrefauthors}%
Hackman, J\BPBI R.%
\BCBT {}\ \BBA {} Katz, N.%
\end{APACrefauthors}%
\unskip\
\newblock
\APACrefYearMonthDay{2010}{}{}.
\newblock
{\BBOQ}\APACrefatitle {Group behavior and performance} {Group behavior and
  performance}.{\BBCQ}
\newblock
\BIn{} S\BPBI T.~Fiske, D\BPBI T.~Gilbert\BCBL {}\ \BBA {} G.~Lindzey\ (\BEDS),
  \APACrefbtitle {Handbook of social psychology} {Handbook of social
  psychology}\ (\BPG~1208–1251).
\newblock
\APACaddressPublisher{}{John Wiley and Sons, Inc.}
\PrintBackRefs{\CurrentBib}

\bibitem [\protect \citeauthoryear {%
Han%
\ \BBA {} Hovav%
}{%
Han%
\ \BBA {} Hovav%
}{%
{\protect \APACyear {2013}}%
}]{%
hanBridgeBondDiverse2013}
\APACinsertmetastar {%
hanBridgeBondDiverse2013}%
\begin{APACrefauthors}%
Han, J.%
\BCBT {}\ \BBA {} Hovav, A.%
\end{APACrefauthors}%
\unskip\
\newblock
\APACrefYearMonthDay{2013}{{\APACmonth{04}}}{}.
\newblock
{\BBOQ}\APACrefatitle {To Bridge or to Bond? {{Diverse}} Social Connections in
  an {{IS}} Project Team} {To bridge or to bond? {{Diverse}} social connections
  in an {{IS}} project team}.{\BBCQ}
\newblock
\APACjournalVolNumPages{International Journal of Project
  Management}{31}{3}{378--390}.
\newblock
\begin{APACrefDOI} \doi{10.1016/j.ijproman.2012.09.001} \end{APACrefDOI}
\PrintBackRefs{\CurrentBib}

\bibitem [\protect \citeauthoryear {%
Hanham%
\ \BBA {} McCormick%
}{%
Hanham%
\ \BBA {} McCormick%
}{%
{\protect \APACyear {2009}}%
}]{%
hanhamGroupWorkSchools2009}
\APACinsertmetastar {%
hanhamGroupWorkSchools2009}%
\begin{APACrefauthors}%
Hanham, J.%
\BCBT {}\ \BBA {} McCormick, J.%
\end{APACrefauthors}%
\unskip\
\newblock
\APACrefYearMonthDay{2009}{{\APACmonth{06}}}{}.
\newblock
{\BBOQ}\APACrefatitle {Group Work in Schools with Close Friends and
  Acquaintances: {{Linking}} Self-Processes with Group Processes} {Group work
  in schools with close friends and acquaintances: {{Linking}} self-processes
  with group processes}.{\BBCQ}
\newblock
\APACjournalVolNumPages{Learning and Instruction}{19}{3}{214--227}.
\newblock
\begin{APACrefDOI} \doi{10.1016/j.learninstruc.2008.04.002} \end{APACrefDOI}
\PrintBackRefs{\CurrentBib}

\bibitem [\protect \citeauthoryear {%
Heinim{\"a}ki%
, Volet%
, Jones%
, Laakkonen%
\BCBL {}\ \BBA {} Vauras%
}{%
Heinim{\"a}ki%
\ \protect \BOthers {.}}{%
{\protect \APACyear {2021}}%
}]{%
heinimakiStudentParticipatoryRole2021a}
\APACinsertmetastar {%
heinimakiStudentParticipatoryRole2021a}%
\begin{APACrefauthors}%
Heinim{\"a}ki, O\BHBI P.%
, Volet, S.%
, Jones, C.%
, Laakkonen, E.%
\BCBL {}\ \BBA {} Vauras, M.%
\end{APACrefauthors}%
\unskip\
\newblock
\APACrefYearMonthDay{2021}{{\APACmonth{09}}}{}.
\newblock
{\BBOQ}\APACrefatitle {Student Participatory Role Profiles in Collaborative
  Science Learning: {{Relation}} of within-Group Configurations of Role
  Profiles and Achievement} {Student participatory role profiles in
  collaborative science learning: {{Relation}} of within-group configurations
  of role profiles and achievement}.{\BBCQ}
\newblock
\APACjournalVolNumPages{Learning, Culture and Social
  Interaction}{30}{}{100539}.
\newblock
\begin{APACrefDOI} \doi{10.1016/j.lcsi.2021.100539} \end{APACrefDOI}
\PrintBackRefs{\CurrentBib}

\bibitem [\protect \citeauthoryear {%
Herbison%
, Benson%
\BCBL {}\ \BBA {} Martin%
}{%
Herbison%
\ \protect \BOthers {.}}{%
{\protect \APACyear {2017}}%
}]{%
herbisonIntricaciesFriendshipcohesionRelationship2017}
\APACinsertmetastar {%
herbisonIntricaciesFriendshipcohesionRelationship2017}%
\begin{APACrefauthors}%
Herbison, J\BPBI D.%
, Benson, A\BPBI J.%
\BCBL {}\ \BBA {} Martin, L\BPBI J.%
\end{APACrefauthors}%
\unskip\
\newblock
\APACrefYearMonthDay{2017}{}{}.
\newblock
{\BBOQ}\APACrefatitle {Intricacies of the Friendship-Cohesion Relationship in
  Children's Sport} {Intricacies of the friendship-cohesion relationship in
  children's sport}.{\BBCQ}
\newblock
\APACjournalVolNumPages{Sport \& Exercise Psychology Review}{13}{1}{10--19}.
\PrintBackRefs{\CurrentBib}

\bibitem [\protect \citeauthoryear {%
Holmes%
, Sholley%
\BCBL {}\ \BBA {} Walker%
}{%
Holmes%
\ \protect \BOthers {.}}{%
{\protect \APACyear {1980}}%
}]{%
holmesLeaderFollowerIsolate1980}
\APACinsertmetastar {%
holmesLeaderFollowerIsolate1980}%
\begin{APACrefauthors}%
Holmes, C\BPBI M.%
, Sholley, B\BPBI K.%
\BCBL {}\ \BBA {} Walker, W\BPBI E.%
\end{APACrefauthors}%
\unskip\
\newblock
\APACrefYearMonthDay{1980}{{\APACmonth{05}}}{}.
\newblock
{\BBOQ}\APACrefatitle {Leader, {{Follower}}, and {{Isolate Personality
  Patterns}} in {{Black}} and {{White Emergent Leadership Groups}}} {Leader,
  {{Follower}}, and {{Isolate Personality Patterns}} in {{Black}} and {{White
  Emergent Leadership Groups}}}.{\BBCQ}
\newblock
\APACjournalVolNumPages{The Journal of Psychology}{105}{1}{41--46}.
\newblock
\begin{APACrefDOI} \doi{10.1080/00223980.1980.9915130} \end{APACrefDOI}
\PrintBackRefs{\CurrentBib}

\bibitem [\protect \citeauthoryear {%
Johnson%
\ \BBA {} Johnson%
}{%
Johnson%
\ \BBA {} Johnson%
}{%
{\protect \APACyear {2008}}%
}]{%
johnsonActiveLearningCooperation2008}
\APACinsertmetastar {%
johnsonActiveLearningCooperation2008}%
\begin{APACrefauthors}%
Johnson%
\BCBT {}\ \BBA {} Johnson, D.%
\end{APACrefauthors}%
\unskip\
\newblock
\APACrefYearMonthDay{2008}{}{}.
\newblock
{\BBOQ}\APACrefatitle {Active {{Learning}}: {{Cooperation}} in the
  {{Classroom}}} {Active {{Learning}}: {{Cooperation}} in the
  {{Classroom}}}.{\BBCQ}
\newblock
\APACjournalVolNumPages{The Annual Report of Educational Psychology in
  Japan}{47}{0}{29--30}.
\newblock
\begin{APACrefDOI} \doi{10.5926/arepj1962.47.0_29} \end{APACrefDOI}
\PrintBackRefs{\CurrentBib}

\bibitem [\protect \citeauthoryear {%
D\BPBI W.~Johnson%
\ \BBA {} Johnson%
}{%
D\BPBI W.~Johnson%
\ \BBA {} Johnson%
}{%
{\protect \APACyear {2002}}%
{\protect \APACexlab {{\protect \BCnt {1}}}}}]{%
johnsonLearningTogetherAlone2002}
\APACinsertmetastar {%
johnsonLearningTogetherAlone2002}%
\begin{APACrefauthors}%
Johnson, D\BPBI W.%
\BCBT {}\ \BBA {} Johnson, R\BPBI T.%
\end{APACrefauthors}%
\unskip\
\newblock
\APACrefYearMonthDay{2002{\protect \BCnt {1}}}{{\APACmonth{01}}}{}.
\newblock
{\BBOQ}\APACrefatitle {Learning {{Together}} and {{Alone}}: {{Overview}} and
  {{Meta}}-analysis} {Learning {{Together}} and {{Alone}}: {{Overview}} and
  {{Meta}}-analysis}.{\BBCQ}
\newblock
\APACjournalVolNumPages{Asia Pacific Journal of Education}{22}{1}{95--105}.
\newblock
\begin{APACrefDOI} \doi{10.1080/0218879020220110} \end{APACrefDOI}
\PrintBackRefs{\CurrentBib}

\bibitem [\protect \citeauthoryear {%
D\BPBI W.~Johnson%
\ \BBA {} Johnson%
}{%
D\BPBI W.~Johnson%
\ \BBA {} Johnson%
}{%
{\protect \APACyear {2002}}%
{\protect \APACexlab {{\protect \BCnt {2}}}}}]{%
johnsonSocialInterdependenceTheory2002}
\APACinsertmetastar {%
johnsonSocialInterdependenceTheory2002}%
\begin{APACrefauthors}%
Johnson, D\BPBI W.%
\BCBT {}\ \BBA {} Johnson, R\BPBI T.%
\end{APACrefauthors}%
\unskip\
\newblock
\APACrefYearMonthDay{2002{\protect \BCnt {2}}}{}{}.
\newblock
{\BBOQ}\APACrefatitle {Social Interdependence Theory and University
  Instruction: {{Theory}} into Practice.} {Social interdependence theory and
  university instruction: {{Theory}} into practice.}{\BBCQ}
\newblock
\APACjournalVolNumPages{Swiss Journal of Psychology/Schweizerische Zeitschrift
  f\"ur Psychologie/Revue Suisse de Psychologie}{61}{3}{119}.
\PrintBackRefs{\CurrentBib}

\bibitem [\protect \citeauthoryear {%
D\BPBI W.~Johnson%
\ \BBA {} Johnson%
}{%
D\BPBI W.~Johnson%
\ \BBA {} Johnson%
}{%
{\protect \APACyear {2009}}%
}]{%
johnsonEducationalPsychologySuccess2009}
\APACinsertmetastar {%
johnsonEducationalPsychologySuccess2009}%
\begin{APACrefauthors}%
Johnson, D\BPBI W.%
\BCBT {}\ \BBA {} Johnson, R\BPBI T.%
\end{APACrefauthors}%
\unskip\
\newblock
\APACrefYearMonthDay{2009}{{\APACmonth{06}}}{}.
\newblock
{\BBOQ}\APACrefatitle {An {{Educational Psychology Success Story}}: {{Social
  Interdependence Theory}} and {{Cooperative Learning}}} {An {{Educational
  Psychology Success Story}}: {{Social Interdependence Theory}} and
  {{Cooperative Learning}}}.{\BBCQ}
\newblock
\APACjournalVolNumPages{Educational Researcher}{38}{5}{365--379}.
\newblock
\begin{APACrefDOI} \doi{10.3102/0013189X09339057} \end{APACrefDOI}
\PrintBackRefs{\CurrentBib}

\bibitem [\protect \citeauthoryear {%
D\BPBI W.~Johnson%
\ \BBA {} Johnson%
}{%
D\BPBI W.~Johnson%
\ \BBA {} Johnson%
}{%
{\protect \APACyear {2014}}%
}]{%
johnsonCooperativeLearning21st2014}
\APACinsertmetastar {%
johnsonCooperativeLearning21st2014}%
\begin{APACrefauthors}%
Johnson, D\BPBI W.%
\BCBT {}\ \BBA {} Johnson, R\BPBI T.%
\end{APACrefauthors}%
\unskip\
\newblock
\APACrefYearMonthDay{2014}{{\APACmonth{10}}}{}.
\newblock
{\BBOQ}\APACrefatitle {Cooperative {{Learning}} in 21st {{Century}}.
  [{{Aprendizaje}} Cooperativo En El Siglo {{XXI}}]} {Cooperative {{Learning}}
  in 21st {{Century}}. [{{Aprendizaje}} cooperativo en el siglo
  {{XXI}}]}.{\BBCQ}
\newblock
\APACjournalVolNumPages{Anales de Psicolog\'ia}{30}{3}{841--851}.
\newblock
\begin{APACrefDOI} \doi{10.6018/analesps.30.3.201241} \end{APACrefDOI}
\PrintBackRefs{\CurrentBib}

\bibitem [\protect \citeauthoryear {%
D\BPBI W.~Johnson%
, Johnson%
\BCBL {}\ \BBA {} Smith%
}{%
D\BPBI W.~Johnson%
\ \protect \BOthers {.}}{%
{\protect \APACyear {2007}}%
}]{%
johnsonStateCooperativeLearning2007a}
\APACinsertmetastar {%
johnsonStateCooperativeLearning2007a}%
\begin{APACrefauthors}%
Johnson, D\BPBI W.%
, Johnson, R\BPBI T.%
\BCBL {}\ \BBA {} Smith, K.%
\end{APACrefauthors}%
\unskip\
\newblock
\APACrefYearMonthDay{2007}{{\APACmonth{02}}}{}.
\newblock
{\BBOQ}\APACrefatitle {The {{State}} of {{Cooperative Learning}} in
  {{Postsecondary}} and {{Professional Settings}}} {The {{State}} of
  {{Cooperative Learning}} in {{Postsecondary}} and {{Professional
  Settings}}}.{\BBCQ}
\newblock
\APACjournalVolNumPages{Educational Psychology Review}{19}{1}{15--29}.
\newblock
\begin{APACrefDOI} \doi{10.1007/s10648-006-9038-8} \end{APACrefDOI}
\PrintBackRefs{\CurrentBib}

\bibitem [\protect \citeauthoryear {%
Landaeta%
\ \protect \BOthers {.}}{%
Landaeta%
\ \protect \BOthers {.}}{%
{\protect \APACyear {2023}}%
}]{%
landaeta2023game}
\APACinsertmetastar {%
landaeta2023game}%
\begin{APACrefauthors}%
Landaeta, V.%
, Candia, C.%
, Pulgar, J.%
, Fabrega, J.%
, Varela, J.%
, Yaikin, T.%
\BDBL {}Rodriguez-Sickert, C.%
\end{APACrefauthors}%
\unskip\
\newblock
\APACrefYearMonthDay{2023}{}{}.
\newblock
{\BBOQ}\APACrefatitle {Game of Tokens: Low Cooperative Relationships as a Key
  Marker for Bully-Victims in Elementary School Classroom} {Game of tokens: Low
  cooperative relationships as a key marker for bully-victims in elementary
  school classroom}.{\BBCQ}
\newblock

\PrintBackRefs{\CurrentBib}

\bibitem [\protect \citeauthoryear {%
Larson%
}{%
Larson%
}{%
{\protect \APACyear {2007}}%
}]{%
larsonDeepDiversityStrong2007}
\APACinsertmetastar {%
larsonDeepDiversityStrong2007}%
\begin{APACrefauthors}%
Larson, J\BPBI R.%
\end{APACrefauthors}%
\unskip\
\newblock
\APACrefYearMonthDay{2007}{{\APACmonth{06}}}{}.
\newblock
{\BBOQ}\APACrefatitle {Deep {{Diversity}} and {{Strong Synergy}}: {{Modeling}}
  the {{Impact}} of {{Variability}} in {{Members}}' {{Problem-Solving
  Strategies}} on {{Group Problem-Solving Performance}}} {Deep {{Diversity}}
  and {{Strong Synergy}}: {{Modeling}} the {{Impact}} of {{Variability}} in
  {{Members}}' {{Problem-Solving Strategies}} on {{Group Problem-Solving
  Performance}}}.{\BBCQ}
\newblock
\APACjournalVolNumPages{Small Group Research}{38}{3}{413--436}.
\newblock
\begin{APACrefDOI} \doi{10.1177/1046496407301972} \end{APACrefDOI}
\PrintBackRefs{\CurrentBib}

\bibitem [\protect \citeauthoryear {%
Laybourn%
, Goldfinch%
, Graham%
, MacLeod%
\BCBL {}\ \BBA {} Stewart%
}{%
Laybourn%
\ \protect \BOthers {.}}{%
{\protect \APACyear {2001}}%
}]{%
laybournMeasuringChangesGroupworking2001}
\APACinsertmetastar {%
laybournMeasuringChangesGroupworking2001}%
\begin{APACrefauthors}%
Laybourn, P.%
, Goldfinch, J.%
, Graham, J.%
, MacLeod, L.%
\BCBL {}\ \BBA {} Stewart, S.%
\end{APACrefauthors}%
\unskip\
\newblock
\APACrefYearMonthDay{2001}{{\APACmonth{08}}}{}.
\newblock
{\BBOQ}\APACrefatitle {Measuring {{Changes}} in {{Groupworking Skills}} in
  {{Undergraduate Students After Employer Involvement}} in {{Group Skill
  Development}}} {Measuring {{Changes}} in {{Groupworking Skills}} in
  {{Undergraduate Students After Employer Involvement}} in {{Group Skill
  Development}}}.{\BBCQ}
\newblock
\APACjournalVolNumPages{Assessment \& Evaluation in Higher
  Education}{26}{4}{367--380}.
\newblock
\begin{APACrefDOI} \doi{10.1080/02602930120063510} \end{APACrefDOI}
\PrintBackRefs{\CurrentBib}

\bibitem [\protect \citeauthoryear {%
Liang%
, Shih%
\BCBL {}\ \BBA {} Chiang%
}{%
Liang%
\ \protect \BOthers {.}}{%
{\protect \APACyear {2015}}%
}]{%
liangTeamDiversityTeam2015}
\APACinsertmetastar {%
liangTeamDiversityTeam2015}%
\begin{APACrefauthors}%
Liang, H\BHBI Y.%
, Shih, H\BHBI A.%
\BCBL {}\ \BBA {} Chiang, Y\BHBI H.%
\end{APACrefauthors}%
\unskip\
\newblock
\APACrefYearMonthDay{2015}{{\APACmonth{02}}}{}.
\newblock
{\BBOQ}\APACrefatitle {Team Diversity and Team Helping Behavior: {{The}}
  Mediating Roles of Team Cooperation and Team Cohesion} {Team diversity and
  team helping behavior: {{The}} mediating roles of team cooperation and team
  cohesion}.{\BBCQ}
\newblock
\APACjournalVolNumPages{European Management Journal}{33}{1}{48--59}.
\newblock
\begin{APACrefDOI} \doi{10.1016/j.emj.2014.07.002} \end{APACrefDOI}
\PrintBackRefs{\CurrentBib}

\bibitem [\protect \citeauthoryear {%
Lopes%
, Mestre%
, Guil%
, Kremenitzer%
\BCBL {}\ \BBA {} Salovey%
}{%
Lopes%
\ \protect \BOthers {.}}{%
{\protect \APACyear {2012}}%
}]{%
lopesRoleKnowledgeSkills2012}
\APACinsertmetastar {%
lopesRoleKnowledgeSkills2012}%
\begin{APACrefauthors}%
Lopes, P\BPBI N.%
, Mestre, J\BPBI M.%
, Guil, R.%
, Kremenitzer, J\BPBI P.%
\BCBL {}\ \BBA {} Salovey, P.%
\end{APACrefauthors}%
\unskip\
\newblock
\APACrefYearMonthDay{2012}{{\APACmonth{08}}}{}.
\newblock
{\BBOQ}\APACrefatitle {The {{Role}} of {{Knowledge}} and {{Skills}} for
  {{Managing Emotions}} in {{Adaptation}} to {{School}}: {{Social Behavior}}
  and {{Misconduct}} in the {{Classroom}}} {The {{Role}} of {{Knowledge}} and
  {{Skills}} for {{Managing Emotions}} in {{Adaptation}} to {{School}}:
  {{Social Behavior}} and {{Misconduct}} in the {{Classroom}}}.{\BBCQ}
\newblock
\APACjournalVolNumPages{American Educational Research
  Journal}{49}{4}{710--742}.
\newblock
\begin{APACrefDOI} \doi{10.3102/0002831212443077} \end{APACrefDOI}
\PrintBackRefs{\CurrentBib}

\bibitem [\protect \citeauthoryear {%
Lou%
, Abrami%
\BCBL {}\ \BBA {} {d'Apollonia}%
}{%
Lou%
\ \protect \BOthers {.}}{%
{\protect \APACyear {2001}}%
}]{%
louSmallGroupIndividual2001a}
\APACinsertmetastar {%
louSmallGroupIndividual2001a}%
\begin{APACrefauthors}%
Lou, Y.%
, Abrami, P\BPBI C.%
\BCBL {}\ \BBA {} {d'Apollonia}, S.%
\end{APACrefauthors}%
\unskip\
\newblock
\APACrefYearMonthDay{2001}{{\APACmonth{09}}}{}.
\newblock
{\BBOQ}\APACrefatitle {Small {{Group}} and {{Individual Learning}} with
  {{Technology}}: {{A Meta-Analysis}}} {Small {{Group}} and {{Individual
  Learning}} with {{Technology}}: {{A Meta-Analysis}}}.{\BBCQ}
\newblock
\APACjournalVolNumPages{Review of Educational Research}{71}{3}{449--521}.
\newblock
\begin{APACrefDOI} \doi{10.3102/00346543071003449} \end{APACrefDOI}
\PrintBackRefs{\CurrentBib}

\bibitem [\protect \citeauthoryear {%
McFarland%
, Moody%
, Diehl%
, Smith%
\BCBL {}\ \BBA {} Thomas%
}{%
McFarland%
\ \protect \BOthers {.}}{%
{\protect \APACyear {2014}}%
}]{%
mcfarlandNetworkEcologyAdolescent2014}
\APACinsertmetastar {%
mcfarlandNetworkEcologyAdolescent2014}%
\begin{APACrefauthors}%
McFarland, D\BPBI A.%
, Moody, J.%
, Diehl, D.%
, Smith, J\BPBI A.%
\BCBL {}\ \BBA {} Thomas, R\BPBI J.%
\end{APACrefauthors}%
\unskip\
\newblock
\APACrefYearMonthDay{2014}{{\APACmonth{12}}}{}.
\newblock
{\BBOQ}\APACrefatitle {Network {{Ecology}} and {{Adolescent Social Structure}}}
  {Network {{Ecology}} and {{Adolescent Social Structure}}}.{\BBCQ}
\newblock
\APACjournalVolNumPages{American Sociological Review}{79}{6}{1088--1121}.
\newblock
\begin{APACrefDOI} \doi{10.1177/0003122414554001} \end{APACrefDOI}
\PrintBackRefs{\CurrentBib}

\bibitem [\protect \citeauthoryear {%
Mercer%
}{%
Mercer%
}{%
{\protect \APACyear {2008}}%
}]{%
mercerTalkDevelopmentReasoning2008}
\APACinsertmetastar {%
mercerTalkDevelopmentReasoning2008}%
\begin{APACrefauthors}%
Mercer, N.%
\end{APACrefauthors}%
\unskip\
\newblock
\APACrefYearMonthDay{2008}{}{}.
\newblock
{\BBOQ}\APACrefatitle {Talk and the {{Development}} of {{Reasoning}} and
  {{Understanding}}} {Talk and the {{Development}} of {{Reasoning}} and
  {{Understanding}}}.{\BBCQ}
\newblock
\APACjournalVolNumPages{Human Development}{51}{1}{90--100}.
\newblock
\begin{APACrefDOI} \doi{10.1159/000113158} \end{APACrefDOI}
\PrintBackRefs{\CurrentBib}

\bibitem [\protect \citeauthoryear {%
Michaelsen%
, Knight%
\BCBL {}\ \BBA {} Fink%
}{%
Michaelsen%
\ \protect \BOthers {.}}{%
{\protect \APACyear {2002}}%
}]{%
michaelsenTeambasedLearningTransformative2002}
\APACinsertmetastar {%
michaelsenTeambasedLearningTransformative2002}%
\begin{APACrefauthors}%
Michaelsen, L\BPBI K.%
, Knight, A\BPBI B.%
\BCBL {}\ \BBA {} Fink, L\BPBI D.%
\end{APACrefauthors}%
\unskip\
\newblock
\APACrefYear{2002}.
\newblock
\APACrefbtitle {Team-Based Learning: {{A}} Transformative Use of Small Groups}
  {Team-based learning: {{A}} transformative use of small groups}.
\newblock
\APACaddressPublisher{}{{Greenwood publishing group}}.
\PrintBackRefs{\CurrentBib}

\bibitem [\protect \citeauthoryear {%
Moradi%
, Faghiharam%
\BCBL {}\ \BBA {} Ghasempour%
}{%
Moradi%
\ \protect \BOthers {.}}{%
{\protect \APACyear {2018}}%
}]{%
moradiRelationshipGroupLearning2018}
\APACinsertmetastar {%
moradiRelationshipGroupLearning2018}%
\begin{APACrefauthors}%
Moradi, S.%
, Faghiharam, B.%
\BCBL {}\ \BBA {} Ghasempour, K.%
\end{APACrefauthors}%
\unskip\
\newblock
\APACrefYearMonthDay{2018}{{\APACmonth{04}}}{}.
\newblock
{\BBOQ}\APACrefatitle {Relationship {{Between Group Learning}} and
  {{Interpersonal Skills With Emphasis}} on the {{Role}} of {{Mediating
  Emotional Intelligence Among High School Students}}} {Relationship {{Between
  Group Learning}} and {{Interpersonal Skills With Emphasis}} on the {{Role}}
  of {{Mediating Emotional Intelligence Among High School Students}}}.{\BBCQ}
\newblock
\APACjournalVolNumPages{SAGE Open}{8}{2}{2158244018782734}.
\newblock
\begin{APACrefDOI} \doi{10.1177/2158244018782734} \end{APACrefDOI}
\PrintBackRefs{\CurrentBib}

\bibitem [\protect \citeauthoryear {%
Myers%
}{%
Myers%
}{%
{\protect \APACyear {2012}}%
}]{%
myersStudentsPerceptionsClassroom2012}
\APACinsertmetastar {%
myersStudentsPerceptionsClassroom2012}%
\begin{APACrefauthors}%
Myers, S\BPBI A.%
\end{APACrefauthors}%
\unskip\
\newblock
\APACrefYearMonthDay{2012}{{\APACmonth{01}}}{}.
\newblock
{\BBOQ}\APACrefatitle {Students' {{Perceptions}} of {{Classroom Group Work}} as
  a {{Function}} of {{Group Member Selection}}} {Students' {{Perceptions}} of
  {{Classroom Group Work}} as a {{Function}} of {{Group Member
  Selection}}}.{\BBCQ}
\newblock
\APACjournalVolNumPages{Communication Teacher}{26}{1}{50--64}.
\newblock
\begin{APACrefDOI} \doi{10.1080/17404622.2011.625368} \end{APACrefDOI}
\PrintBackRefs{\CurrentBib}

\bibitem [\protect \citeauthoryear {%
Oakley%
, Felder%
, Brent%
\BCBL {}\ \BBA {} Elhajj%
}{%
Oakley%
\ \protect \BOthers {.}}{%
{\protect \APACyear {2004}}%
}]{%
oakleyTurningStudentGroups}
\APACinsertmetastar {%
oakleyTurningStudentGroups}%
\begin{APACrefauthors}%
Oakley, B.%
, Felder, R\BPBI M.%
, Brent, R.%
\BCBL {}\ \BBA {} Elhajj, I.%
\end{APACrefauthors}%
\unskip\
\newblock
\APACrefYearMonthDay{2004}{}{}.
\newblock
{\BBOQ}\APACrefatitle {Turning Student Groups into Effective Teams} {Turning
  student groups into effective teams}.{\BBCQ}
\newblock
\APACjournalVolNumPages{Journal of student centered learning}{2}{1}{9--34}.
\PrintBackRefs{\CurrentBib}

\bibitem [\protect \citeauthoryear {%
Osterman%
}{%
Osterman%
}{%
{\protect \APACyear {2000}}%
}]{%
ostermanStudentsNeedBelonging2000}
\APACinsertmetastar {%
ostermanStudentsNeedBelonging2000}%
\begin{APACrefauthors}%
Osterman, K\BPBI F.%
\end{APACrefauthors}%
\unskip\
\newblock
\APACrefYearMonthDay{2000}{{\APACmonth{09}}}{}.
\newblock
{\BBOQ}\APACrefatitle {Students' {{Need}} for {{Belonging}} in the {{School
  Community}}} {Students' {{Need}} for {{Belonging}} in the {{School
  Community}}}.{\BBCQ}
\newblock
\APACjournalVolNumPages{Review of Educational Research}{70}{3}{323--367}.
\newblock
\begin{APACrefDOI} \doi{10.3102/00346543070003323} \end{APACrefDOI}
\PrintBackRefs{\CurrentBib}

\bibitem [\protect \citeauthoryear {%
Park%
, Kim%
\BCBL {}\ \BBA {} Gully%
}{%
Park%
\ \protect \BOthers {.}}{%
{\protect \APACyear {2017}}%
}]{%
parkEffectCohesionCurvilinear2017}
\APACinsertmetastar {%
parkEffectCohesionCurvilinear2017}%
\begin{APACrefauthors}%
Park, W\BHBI W.%
, Kim, M\BPBI S.%
\BCBL {}\ \BBA {} Gully, S\BPBI M.%
\end{APACrefauthors}%
\unskip\
\newblock
\APACrefYearMonthDay{2017}{{\APACmonth{08}}}{}.
\newblock
{\BBOQ}\APACrefatitle {Effect of {{Cohesion}} on the {{Curvilinear Relationship
  Between Team Efficacy}} and {{Performance}}} {Effect of {{Cohesion}} on the
  {{Curvilinear Relationship Between Team Efficacy}} and
  {{Performance}}}.{\BBCQ}
\newblock
\APACjournalVolNumPages{Small Group Research}{48}{4}{455--481}.
\newblock
\begin{APACrefDOI} \doi{10.1177/1046496417709933} \end{APACrefDOI}
\PrintBackRefs{\CurrentBib}

\bibitem [\protect \citeauthoryear {%
Pulgar%
, Candia%
\BCBL {}\ \BBA {} Leonardi%
}{%
Pulgar%
\ \protect \BOthers {.}}{%
{\protect \APACyear {2020}}%
}]{%
pulgar_etal2020}
\APACinsertmetastar {%
pulgar_etal2020}%
\begin{APACrefauthors}%
Pulgar, J.%
, Candia, C.%
\BCBL {}\ \BBA {} Leonardi, P\BPBI M.%
\end{APACrefauthors}%
\unskip\
\newblock
\APACrefYearMonthDay{2020}{Jun}{}.
\newblock
{\BBOQ}\APACrefatitle {Social networks and academic performance in physics:
  Undergraduate cooperation enhances ill-structured problem elaboration and
  inhibits well-structured problem solving} {Social networks and academic
  performance in physics: Undergraduate cooperation enhances ill-structured
  problem elaboration and inhibits well-structured problem solving}.{\BBCQ}
\newblock
\APACjournalVolNumPages{Phys. Rev. Phys. Educ. Res.}{16}{}{010137}.
\PrintBackRefs{\CurrentBib}

\bibitem [\protect \citeauthoryear {%
Pulgar%
, Ram\'{\i}rez%
, Umanzor%
, Candia%
\BCBL {}\ \BBA {} S\'anchez%
}{%
Pulgar%
\ \protect \BOthers {.}}{%
{\protect \APACyear {2022}}%
}]{%
pulgar_etal_2022}
\APACinsertmetastar {%
pulgar_etal_2022}%
\begin{APACrefauthors}%
Pulgar, J.%
, Ram\'{\i}rez, D.%
, Umanzor, A.%
, Candia, C.%
\BCBL {}\ \BBA {} S\'anchez, I.%
\end{APACrefauthors}%
\unskip\
\newblock
\APACrefYearMonthDay{2022}{Jun}{}.
\newblock
{\BBOQ}\APACrefatitle {Long-term collaboration with strong friendship ties
  improves academic performance in remote and hybrid teaching modalities in
  high school physics} {Long-term collaboration with strong friendship ties
  improves academic performance in remote and hybrid teaching modalities in
  high school physics}.{\BBCQ}
\newblock
\APACjournalVolNumPages{Phys. Rev. Phys. Educ. Res.}{18}{}{010146}.
\newblock
\begin{APACrefURL}
  \url{https://link.aps.org/doi/10.1103/PhysRevPhysEducRes.18.010146}
  \end{APACrefURL}
\newblock
\begin{APACrefDOI} \doi{10.1103/PhysRevPhysEducRes.18.010146} \end{APACrefDOI}
\PrintBackRefs{\CurrentBib}

\bibitem [\protect \citeauthoryear {%
Puurtinen%
, Heap%
\BCBL {}\ \BBA {} Mappes%
}{%
Puurtinen%
\ \protect \BOthers {.}}{%
{\protect \APACyear {2015}}%
}]{%
puurtinenJointEmergenceGroup2015}
\APACinsertmetastar {%
puurtinenJointEmergenceGroup2015}%
\begin{APACrefauthors}%
Puurtinen, M.%
, Heap, S.%
\BCBL {}\ \BBA {} Mappes, T.%
\end{APACrefauthors}%
\unskip\
\newblock
\APACrefYearMonthDay{2015}{{\APACmonth{05}}}{}.
\newblock
{\BBOQ}\APACrefatitle {The Joint Emergence of Group Competition and
  Within-Group Cooperation} {The joint emergence of group competition and
  within-group cooperation}.{\BBCQ}
\newblock
\APACjournalVolNumPages{Evolution and Human Behavior}{36}{3}{211--217}.
\newblock
\begin{APACrefDOI} \doi{10.1016/j.evolhumbehav.2014.11.005} \end{APACrefDOI}
\PrintBackRefs{\CurrentBib}

\bibitem [\protect \citeauthoryear {%
{R Core Team}%
}{%
{R Core Team}%
}{%
{\protect \APACyear {2020}}%
}]{%
rstudio}
\APACinsertmetastar {%
rstudio}%
\begin{APACrefauthors}%
{R Core Team}.%
\end{APACrefauthors}%
\unskip\
\newblock
\APACrefYearMonthDay{2020}{}{}.
\newblock
{\BBOQ}\APACrefatitle {R: A Language and Environment for Statistical Computing}
  {R: A language and environment for statistical computing}{\BBCQ}\
  [\bibcomputersoftwaremanual].
\newblock
\APACaddressPublisher{Vienna, Austria}{}.
\newblock
\begin{APACrefURL} \url{https://www.R-project.org/} \end{APACrefURL}
\PrintBackRefs{\CurrentBib}

\bibitem [\protect \citeauthoryear {%
Rambaran%
\ \protect \BOthers {.}}{%
Rambaran%
\ \protect \BOthers {.}}{%
{\protect \APACyear {2017}}%
}]{%
rambaranAcademicFunctioningPeer2017}
\APACinsertmetastar {%
rambaranAcademicFunctioningPeer2017}%
\begin{APACrefauthors}%
Rambaran, J\BPBI A.%
, Hopmeyer, A.%
, Schwartz, D.%
, Steglich, C.%
, Badaly, D.%
\BCBL {}\ \BBA {} Veenstra, R.%
\end{APACrefauthors}%
\unskip\
\newblock
\APACrefYearMonthDay{2017}{{\APACmonth{03}}}{}.
\newblock
{\BBOQ}\APACrefatitle {Academic {{Functioning}} and {{Peer Influences}}: {{A
  Short-Term Longitudinal Study}} of {{Network-Behavior Dynamics}} in {{Middle
  Adolescence}}} {Academic {{Functioning}} and {{Peer Influences}}: {{A
  Short-Term Longitudinal Study}} of {{Network-Behavior Dynamics}} in {{Middle
  Adolescence}}}.{\BBCQ}
\newblock
\APACjournalVolNumPages{Child Development}{88}{2}{523--543}.
\newblock
\begin{APACrefDOI} \doi{10.1111/cdev.12611} \end{APACrefDOI}
\PrintBackRefs{\CurrentBib}

\bibitem [\protect \citeauthoryear {%
Slavin%
}{%
Slavin%
}{%
{\protect \APACyear {1996}}%
}]{%
slavinResearchCooperativeLearning1996}
\APACinsertmetastar {%
slavinResearchCooperativeLearning1996}%
\begin{APACrefauthors}%
Slavin, R\BPBI E.%
\end{APACrefauthors}%
\unskip\
\newblock
\APACrefYearMonthDay{1996}{{\APACmonth{01}}}{}.
\newblock
{\BBOQ}\APACrefatitle {Research on {{Cooperative Learning}} and
  {{Achievement}}: {{What We Know}}, {{What We Need}} to {{Know}}} {Research on
  {{Cooperative Learning}} and {{Achievement}}: {{What We Know}}, {{What We
  Need}} to {{Know}}}.{\BBCQ}
\newblock
\APACjournalVolNumPages{Contemporary Educational Psychology}{21}{1}{43--69}.
\newblock
\begin{APACrefDOI} \doi{10.1006/ceps.1996.0004} \end{APACrefDOI}
\PrintBackRefs{\CurrentBib}

\bibitem [\protect \citeauthoryear {%
Springer%
, Stanne%
\BCBL {}\ \BBA {} Donovan%
}{%
Springer%
\ \protect \BOthers {.}}{%
{\protect \APACyear {1999}}%
}]{%
springerEffectsSmallGroupLearning1999}
\APACinsertmetastar {%
springerEffectsSmallGroupLearning1999}%
\begin{APACrefauthors}%
Springer, L.%
, Stanne, M\BPBI E.%
\BCBL {}\ \BBA {} Donovan, S\BPBI S.%
\end{APACrefauthors}%
\unskip\
\newblock
\APACrefYearMonthDay{1999}{{\APACmonth{03}}}{}.
\newblock
{\BBOQ}\APACrefatitle {Effects of {{Small-Group Learning}} on
  {{Undergraduates}} in {{Science}}, {{Mathematics}}, {{Engineering}}, and
  {{Technology}}: {{A Meta-Analysis}}} {Effects of {{Small-Group Learning}} on
  {{Undergraduates}} in {{Science}}, {{Mathematics}}, {{Engineering}}, and
  {{Technology}}: {{A Meta-Analysis}}}.{\BBCQ}
\newblock
\APACjournalVolNumPages{Review of Educational Research}{69}{1}{21--51}.
\newblock
\begin{APACrefDOI} \doi{10.3102/00346543069001021} \end{APACrefDOI}
\PrintBackRefs{\CurrentBib}

\bibitem [\protect \citeauthoryear {%
Srba%
\ \BBA {} Bielikova%
}{%
Srba%
\ \BBA {} Bielikova%
}{%
{\protect \APACyear {2015}}%
}]{%
srbaDynamicGroupFormation2015a}
\APACinsertmetastar {%
srbaDynamicGroupFormation2015a}%
\begin{APACrefauthors}%
Srba, I.%
\BCBT {}\ \BBA {} Bielikova, M.%
\end{APACrefauthors}%
\unskip\
\newblock
\APACrefYearMonthDay{2015}{{\APACmonth{04}}}{}.
\newblock
{\BBOQ}\APACrefatitle {Dynamic {{Group Formation}} as an {{Approach}} to
  {{Collaborative Learning Support}}} {Dynamic {{Group Formation}} as an
  {{Approach}} to {{Collaborative Learning Support}}}.{\BBCQ}
\newblock
\APACjournalVolNumPages{IEEE Transactions on Learning
  Technologies}{8}{2}{173--186}.
\newblock
\begin{APACrefDOI} \doi{10.1109/TLT.2014.2373374} \end{APACrefDOI}
\PrintBackRefs{\CurrentBib}

\bibitem [\protect \citeauthoryear {%
Steiner%
}{%
Steiner%
}{%
{\protect \APACyear {1966}}%
}]{%
Steiner1966}
\APACinsertmetastar {%
Steiner1966}%
\begin{APACrefauthors}%
Steiner, I\BPBI D.%
\end{APACrefauthors}%
\unskip\
\newblock
\APACrefYearMonthDay{1966}{}{}.
\newblock
{\BBOQ}\APACrefatitle {Models for inferring relationships between group size
  and potential productivity} {Models for inferring relationships between group
  size and potential productivity}.{\BBCQ}
\newblock
\APACjournalVolNumPages{Behavioral Science}{11}{}{273--283}.
\PrintBackRefs{\CurrentBib}

\bibitem [\protect \citeauthoryear {%
Steiner%
}{%
Steiner%
}{%
{\protect \APACyear {1972}}%
}]{%
steinerGroupProcessProductivity1972}
\APACinsertmetastar {%
steinerGroupProcessProductivity1972}%
\begin{APACrefauthors}%
Steiner, I\BPBI D.%
\end{APACrefauthors}%
\unskip\
\newblock
\APACrefYear{1972}.
\newblock
\APACrefbtitle {Group Process and Productivity} {Group process and
  productivity}.
\newblock
\APACaddressPublisher{}{{Academic press New York}}.
\PrintBackRefs{\CurrentBib}

\bibitem [\protect \citeauthoryear {%
Tuckman%
}{%
Tuckman%
}{%
{\protect \APACyear {1965}}%
}]{%
tuckmanDevelopmentalSequenceSmall1965}
\APACinsertmetastar {%
tuckmanDevelopmentalSequenceSmall1965}%
\begin{APACrefauthors}%
Tuckman, B\BPBI W.%
\end{APACrefauthors}%
\unskip\
\newblock
\APACrefYearMonthDay{1965}{{\APACmonth{06}}}{}.
\newblock
{\BBOQ}\APACrefatitle {Developmental Sequence in Small Groups.} {Developmental
  sequence in small groups.}{\BBCQ}
\newblock
\APACjournalVolNumPages{Psychological Bulletin}{63}{6}{384--399}.
\newblock
\begin{APACrefDOI} \doi{10.1037/h0022100} \end{APACrefDOI}
\PrintBackRefs{\CurrentBib}

\bibitem [\protect \citeauthoryear {%
Urhahne%
\ \BBA {} Wijnia%
}{%
Urhahne%
\ \BBA {} Wijnia%
}{%
{\protect \APACyear {2021}}%
}]{%
urhahneReviewAccuracyTeacher2021}
\APACinsertmetastar {%
urhahneReviewAccuracyTeacher2021}%
\begin{APACrefauthors}%
Urhahne, D.%
\BCBT {}\ \BBA {} Wijnia, L.%
\end{APACrefauthors}%
\unskip\
\newblock
\APACrefYearMonthDay{2021}{{\APACmonth{02}}}{}.
\newblock
{\BBOQ}\APACrefatitle {A Review on the Accuracy of Teacher Judgments} {A review
  on the accuracy of teacher judgments}.{\BBCQ}
\newblock
\APACjournalVolNumPages{Educational Research Review}{32}{}{100374}.
\newblock
\begin{APACrefDOI} \doi{10.1016/j.edurev.2020.100374} \end{APACrefDOI}
\PrintBackRefs{\CurrentBib}

\bibitem [\protect \citeauthoryear {%
Vesely%
, Kl{\"o}ckner%
\BCBL {}\ \BBA {} Brick%
}{%
Vesely%
\ \protect \BOthers {.}}{%
{\protect \APACyear {2020}}%
}]{%
veselyProenvironmentalBehaviorSignal2020}
\APACinsertmetastar {%
veselyProenvironmentalBehaviorSignal2020}%
\begin{APACrefauthors}%
Vesely, S.%
, Kl{\"o}ckner, C\BPBI A.%
\BCBL {}\ \BBA {} Brick, C.%
\end{APACrefauthors}%
\unskip\
\newblock
\APACrefYearMonthDay{2020}{{\APACmonth{02}}}{}.
\newblock
{\BBOQ}\APACrefatitle {Pro-Environmental Behavior as a Signal of
  Cooperativeness: {{Evidence}} from a Social Dilemma Experiment}
  {Pro-environmental behavior as a signal of cooperativeness: {{Evidence}} from
  a social dilemma experiment}.{\BBCQ}
\newblock
\APACjournalVolNumPages{Journal of Environmental Psychology}{67}{}{101362}.
\newblock
\begin{APACrefDOI} \doi{10.1016/j.jenvp.2019.101362} \end{APACrefDOI}
\PrintBackRefs{\CurrentBib}

\bibitem [\protect \citeauthoryear {%
V{\"o}llinger%
, Supanc%
\BCBL {}\ \BBA {} Brunstein%
}{%
V{\"o}llinger%
\ \protect \BOthers {.}}{%
{\protect \APACyear {2022}}%
}]{%
vollingerVideobasedStudyStudent2022}
\APACinsertmetastar {%
vollingerVideobasedStudyStudent2022}%
\begin{APACrefauthors}%
V{\"o}llinger, V\BPBI A.%
, Supanc, M.%
\BCBL {}\ \BBA {} Brunstein, J\BPBI C.%
\end{APACrefauthors}%
\unskip\
\newblock
\APACrefYearMonthDay{2022}{{\APACmonth{02}}}{}.
\newblock
{\BBOQ}\APACrefatitle {A Video-Based Study of Student Teachers' Participation
  and Content Processing in Cooperative Group Work} {A video-based study of
  student teachers' participation and content processing in cooperative group
  work}.{\BBCQ}
\newblock
\APACjournalVolNumPages{Learning, Culture and Social
  Interaction}{32}{}{100598}.
\newblock
\begin{APACrefDOI} \doi{10.1016/j.lcsi.2021.100598} \end{APACrefDOI}
\PrintBackRefs{\CurrentBib}

\bibitem [\protect \citeauthoryear {%
Vygotsky%
\ \BBA {} Cole%
}{%
Vygotsky%
\ \BBA {} Cole%
}{%
{\protect \APACyear {1978}}%
}]{%
vygotskyMindSocietyDevelopment1978a}
\APACinsertmetastar {%
vygotskyMindSocietyDevelopment1978a}%
\begin{APACrefauthors}%
Vygotsky, L\BPBI S.%
\BCBT {}\ \BBA {} Cole, M.%
\end{APACrefauthors}%
\unskip\
\newblock
\APACrefYear{1978}.
\newblock
\APACrefbtitle {Mind in Society: {{Development}} of Higher Psychological
  Processes} {Mind in society: {{Development}} of higher psychological
  processes}.
\newblock
\APACaddressPublisher{}{{Harvard university press}}.
\PrintBackRefs{\CurrentBib}

\bibitem [\protect \citeauthoryear {%
Wang%
, {Rubie-Davies}%
\BCBL {}\ \BBA {} Meissel%
}{%
Wang%
\ \protect \BOthers {.}}{%
{\protect \APACyear {2018}}%
}]{%
wangSystematicReviewTeacher2018}
\APACinsertmetastar {%
wangSystematicReviewTeacher2018}%
\begin{APACrefauthors}%
Wang, S.%
, {Rubie-Davies}, C\BPBI M.%
\BCBL {}\ \BBA {} Meissel, K.%
\end{APACrefauthors}%
\unskip\
\newblock
\APACrefYearMonthDay{2018}{{\APACmonth{04}}}{}.
\newblock
{\BBOQ}\APACrefatitle {A Systematic Review of the Teacher Expectation
  Literature over the Past 30 Years} {A systematic review of the teacher
  expectation literature over the past 30 years}.{\BBCQ}
\newblock
\APACjournalVolNumPages{Educational Research and
  Evaluation}{24}{3-5}{124--179}.
\newblock
\begin{APACrefDOI} \doi{10.1080/13803611.2018.1548798} \end{APACrefDOI}
\PrintBackRefs{\CurrentBib}

\bibitem [\protect \citeauthoryear {%
Webb%
, Nemer%
\BCBL {}\ \BBA {} Zuniga%
}{%
Webb%
\ \protect \BOthers {.}}{%
{\protect \APACyear {2002}}%
}]{%
webbShortCircuitsSuperconductors2002}
\APACinsertmetastar {%
webbShortCircuitsSuperconductors2002}%
\begin{APACrefauthors}%
Webb, N\BPBI M.%
, Nemer, K\BPBI M.%
\BCBL {}\ \BBA {} Zuniga, S.%
\end{APACrefauthors}%
\unskip\
\newblock
\APACrefYearMonthDay{2002}{{\APACmonth{01}}}{}.
\newblock
{\BBOQ}\APACrefatitle {Short {{Circuits}} or {{Superconductors}}? {{Effects}}
  of {{Group Composition}} on {{High-Achieving Students}}' {{Science Assessment
  Performance}}} {Short {{Circuits}} or {{Superconductors}}? {{Effects}} of
  {{Group Composition}} on {{High-Achieving Students}}' {{Science Assessment
  Performance}}}.{\BBCQ}
\newblock
\APACjournalVolNumPages{American Educational Research
  Journal}{39}{4}{943--989}.
\newblock
\begin{APACrefDOI} \doi{10.3102/00028312039004943} \end{APACrefDOI}
\PrintBackRefs{\CurrentBib}

\bibitem [\protect \citeauthoryear {%
Wegerif%
}{%
Wegerif%
}{%
{\protect \APACyear {2007}}%
}]{%
wegerifDialogicEducationTechnology2007}
\APACinsertmetastar {%
wegerifDialogicEducationTechnology2007}%
\begin{APACrefauthors}%
Wegerif, R.%
\end{APACrefauthors}%
\unskip\
\newblock
\APACrefYear{2007}.
\newblock
\APACrefbtitle {Dialogic Education and Technology: {{Expanding}} the Space of
  Learning} {Dialogic education and technology: {{Expanding}} the space of
  learning}\ (\BVOL~7).
\newblock
\APACaddressPublisher{}{{Springer Science \& Business Media}}.
\PrintBackRefs{\CurrentBib}

\bibitem [\protect \citeauthoryear {%
Wentzel%
}{%
Wentzel%
}{%
{\protect \APACyear {2017}}%
}]{%
wentzelPeerRelationshipsMotivation2017}
\APACinsertmetastar {%
wentzelPeerRelationshipsMotivation2017}%
\begin{APACrefauthors}%
Wentzel, K\BPBI R.%
\end{APACrefauthors}%
\unskip\
\newblock
\APACrefYearMonthDay{2017}{}{}.
\newblock
{\BBOQ}\APACrefatitle {Peer Relationships, Motivation, and Academic Performance
  at School.} {Peer relationships, motivation, and academic performance at
  school.}{\BBCQ}
\newblock
\BIn{} \APACrefbtitle {Handbook of Competence and Motivation: {{Theory}} and
  Application, 2nd Ed.} {Handbook of competence and motivation: {{Theory}} and
  application, 2nd ed.}\ (\BPGS\ 586--603).
\newblock
\APACaddressPublisher{{New York, NY, US}}{{The Guilford Press}}.
\PrintBackRefs{\CurrentBib}

\bibitem [\protect \citeauthoryear {%
Wichmann%
\ \protect \BOthers {.}}{%
Wichmann%
\ \protect \BOthers {.}}{%
{\protect \APACyear {2016}}%
}]{%
wichmannGroupFormationSmallGroup2016}
\APACinsertmetastar {%
wichmannGroupFormationSmallGroup2016}%
\begin{APACrefauthors}%
Wichmann, A.%
, Hecking, T.%
, Elson, M.%
, Christmann, N.%
, Herrmann, T.%
\BCBL {}\ \BBA {} Hoppe, H\BPBI U.%
\end{APACrefauthors}%
\unskip\
\newblock
\APACrefYearMonthDay{2016}{{\APACmonth{08}}}{}.
\newblock
{\BBOQ}\APACrefatitle {Group {{Formation}} for {{Small-Group Learning}}: {{Are
  Heterogeneous Groups More Productive}}?} {Group {{Formation}} for
  {{Small-Group Learning}}: {{Are Heterogeneous Groups More
  Productive}}?}{\BBCQ}
\newblock
\BIn{} \APACrefbtitle {Proceedings of the 12th {{International Symposium}} on
  {{Open Collaboration}}} {Proceedings of the 12th {{International Symposium}}
  on {{Open Collaboration}}}\ (\BPGS\ 1--4).
\newblock
\APACaddressPublisher{{Berlin Germany}}{{ACM}}.
\newblock
\begin{APACrefDOI} \doi{10.1145/2957792.2965662} \end{APACrefDOI}
\PrintBackRefs{\CurrentBib}

\bibitem [\protect \citeauthoryear {%
Zins%
}{%
Zins%
}{%
{\protect \APACyear {2004}}%
}]{%
zinsBuildingAcademicSuccess2004}
\APACinsertmetastar {%
zinsBuildingAcademicSuccess2004}%
\begin{APACrefauthors}%
Zins, J\BPBI E.%
\end{APACrefauthors}%
\unskip\
\newblock
\APACrefYear{2004}.
\newblock
\APACrefbtitle {Building {{Academic Success}} on {{Social}} and {{Emotional
  Learning}}: {{What Does}} the {{Research Say}}?} {Building {{Academic
  Success}} on {{Social}} and {{Emotional Learning}}: {{What Does}} the
  {{Research Say}}?}
\newblock
\APACaddressPublisher{}{{Teachers College Press}}.
\PrintBackRefs{\CurrentBib}

\end{thebibliography}

\newpage

\section{Supplementary
material}

In the following section, we show the results related to the composition of the groups, in addition to providing supplementary material associated with the models presented in this work.

\begin{table*}[!h]
\begin{center}
\caption{Descriptive statistics of directed networks at the end of Semester 1: Friendship ties, Physics prestige nomination, and Physics collaboration.
\label{tab:parametrosred2}}
  \centering
 \begin{tabular}{lcccc|ccc}
\hline
       &        & \multicolumn{3}{c}{Friendship ($t_0$)} & \multicolumn{3}{c}{Friendship ($t_1$)} \\
\cline{2-8}
\\
\multicolumn{1}{c}{Course} &\multicolumn{1}{c}{ Nodes}  & \multicolumn{1}{c}{Links} & \multicolumn{1}{c}{Degree$^*$}& \multicolumn{1}{c}{Density}   & \multicolumn{1}{c}{Links} & \multicolumn{1}{c}{Degree$^*$} &\multicolumn{1}{c}{ Density} 
\\
\cline{1-8}
\\
E1     &	24	&	175	&	7.29	&	0.32	&	151	&	6.29	&	0.27  \\
E2     &	22	&	125	&	5.68	&	0.27	&	118	&	5.36	&	0.26 \\ 
C1     &	24	&	117	&	4.88	&	0.21	&	112	&	4.67	&	0.20 \\
C2     &	20	&	117	&	5.85	&	0.31	&	95	&	4.75	&	0.25 \\
\\
 \hline
  \\
       &        &  \multicolumn{3}{c}{Prestige in Physics} & \multicolumn{3}{c}{Collaboration in Physics} \\
\cline{2-8}
\\
\multicolumn{1}{c}{Course} &\multicolumn{1}{c}{ Nodes}  & \multicolumn{1}{c}{Links} & \multicolumn{1}{c}{Degree$^*$}& \multicolumn{1}{c}{Density}   & \multicolumn{1}{c}{Links} & \multicolumn{1}{c}{Degree$^*$} &\multicolumn{1}{c}{ Density} 
\\
\cline{1-8}
\\
E1     &	24	&	127	&	5.29	&	0.23	&	113	&	4.71	&	0.21 \\
E2     &	22	&	108	&	4.91	&	0.23	&	71	&	3.22	&	0.15 \\ 
C1     &	24	&	147	&	6.13	&	0.27	&	134	&	5.58	&	0.24 \\
C2     &	20	&	101	&	5.05	&	0.27	&	57	&	2.85	&	0.15 \\
\\
\hline
\multicolumn{8}{l}{$^*$Average number of incoming and outgoing links per students in the class.} 
\end{tabular}
\end{center}
\end{table*}

\begin{table}[!htbp] \centering 
  \caption{Regression models for the network variables friendship, cooperation and prestige outside the work group. Unlike the models presented in the main paper, the treatment used does not generate any effect on the predicted network variables.} 
  \label{tab:s2} 
   \resizebox{1\textwidth}{!}{ 
\begin{tabular}{lcccc|cccc} 
\cline{1-9}\\
 & \multicolumn{8}{c}{\textit{Dependent variable:}} \\ 
[0.5ex]\cline{2-9}
\\ &  \multicolumn{4}{c}{Friendship outside Team ($t_1$)} & \multicolumn{2}{c}{ Cooperation outside Team ($t_1$)} & \multicolumn{2}{c}{Prestige outside Team ($t_1$) }\\
\\ &  \multicolumn{4}{c}{(in-degree)} & \multicolumn{2}{c}{(in-degree)} & \multicolumn{2}{c}{(in-degree) }\\
\\ 
[0.5ex]\cline{2-9}
	\\&(1)&(2)&(3)&(4)&(5)&(6)&(7)&(8)	\\
 \hline \\ 

Exp. (Afinity-based groups)	&	0.39	&	-0.32	&	-0.69$^{\dagger}$	&	-1.00	&	0.04	&	0.41	&	0.10	&	-0.40	\\
	&	(0.36)	&	(0.69)	&	(0.36)	&	(0.75)	&	(0.22)	&	(0.42)	&	(0.79)	&	(1.53)	\\
 Group Size (3)	&	0.91$^{\dagger}$	&	0.88$^{\dagger}$	&	1.12$^{*}$	&	1.12$^{*}$	&	0.53$^{\dagger}$	&	0.55$^{\dagger}$	&	-0.12	&	-0.14	\\
	&	(0.52)	&	(0.52)	&	(0.51)	&	(0.51)	&	(0.31)	&	(0.31)	&	(1.15)	&	(1.16)	\\
 Gender (Female)	&	-0.14	&	-0.17	&	-0.02	&	-0.03	&	-0.15	&	-0.13	&	-2.65$^{***}$	&	-2.68$^{***}$	\\
	&	(0.35)	&	(0.35)	&	(0.34)	&	(0.34)	&	(0.21)	&	(0.21)	&	(0.77)	&	(0.78)	\\
Grade Level (10th)	&	-0.60	&	-0.54	&	-1.02$^{*}$	&	-1.00$^{*}$	&	-2.39$^{***}$	&	-2.42$^{***}$	&	-0.80	&	-0.76	\\
	&	(0.44)	&	(0.44)	&	(0.43)	&	(0.43)	&	(0.27)	&	(0.27)	&	(0.98)	&	(0.99)	\\
Average grades	&	0.06$^{*}$	&	0.06$^{*}$	&	0.06*	&	0.06$^{*}$	&	-0.01	&	-0.01	&	0.19$^{**}$	&	0.19$^{**}$	\\
	&	(0.03)	&	(0.03)	&	(0.03)	&	(0.03)	&	(0.02)	&	(0.02)	&	(0.06)	&	(0.06)	\\
Friendship outside team ($t_0$)	&	0.84$^{***}$	&	0.72$^{***}	$&		&		&	0.11$^{*}$	&	0.17$^{*}$	&	0.61$^{***}$	&	0.53$^{\dagger}$	\\
(in-degree)	&	(0.07)	&	(0.12)	&		&		&	(0.04)	&	(0.07)	&	(0.16)	&	(0.27)	\\
Exp.$*$ Friendship outside team ($t_0$)	&		&	0.17	&		&		&		&	-0.09	&		&	0.13	\\
(in-degree)	&		&	(0.15)	&		&		&		&	(0.09)	&		&	(0.32)	\\
Friendship ($t_0$)	&		&		&	0.74$^{***}$	&	0.70$^{***}$	&		&		&		&		\\
(in-degree)	&		&		&	(0.06)	&	(0.11)	&		&		&		&		\\
Exp.$*$Friendship ($t_0$)	&		&		&		&	0.06	&		&		&		&		\\
(in-degree)	&		&		&		&	(0.13)	&		&		&		&		\\
Intercept	&	-2.49	&	-1.98	&	-2.46	&	-2.25	&	3.22$^{**}$	&	2.96$^{**}	$&	-8.01$^{*}$	&	-7.64$^{\dagger}	$\\
	&	(1.72)	&	(1.77)	&	(1.69)	&	(1.76)	&	(1.04)	&	(1.07)	&	(3.82)	&	(3.96)	\\

 \hline \\ 	
Num.Obs.	&	90	&	90	&	90	&	90	&	90	&	90	&	90	&	90	\\
R2	&	0.67	&	0.67	&	0.68	&	0.68	&	0.62	&	0.63	&	0.33	&	0.34	\\
R2 Adj.	&	0.64	&	0.64	&	0.66	&	0.65	&	0.59	&	0.59	&	0.29	&	0.28	\\
AIC	&	349.3	&	349.7	&	345.8	&	347.5	&	258.5	&	259.4	&	492.7	&	494.5	\\
BIC	&	369.3	&	372.2	&	365.8	&	370.0	&	278.5	&	281.9	&	512.6	&	517.0	\\
Log.Lik.	&	-166.65	&	-165.86	&	-164.89	&	-164.77	&	-121.25	&	-120.69	&	-238.32	&	-238.24	\\
F	&	27.73	&	24.10	&	29.38	&	24.98	&	22.73	&	19.64	&	6.96	&	5.92	\\
RMSE	&	1.61	&	1.60	&	1.57	&	1.58	&	0.97	&	0.97	&	3.56	&	3.58	\\

\\
\hline 

\textit{Note:}  & \multicolumn{8}{r}{$^{\dagger}$ p $<$ 0.1, $^{*}$ p $<$ 0.05, $^{**}$ p $<$ 0.01, $^{***}$p $<$ 0.001} \\ 
\end{tabular} 
}
\end{table}

\begin{table}[!htbp] \centering 
  \caption{The variance inflation factor (VIF) statistics for the predictor and control variables in the regression models for n students with whom a person continues to work, presented in the manuscript.} 
  \label{tab:s3} 
   \resizebox{1\textwidth}{!}{ 
\begin{tabular}{lcccc|cc|cc} 
\cline{1-9}\\
 & \multicolumn{8}{c}{\textit{Dependent variable:}} \\ 
[0.5ex]\cline{2-9}
\\ &  \multicolumn{4}{c}{Friendship outside Team ($t_1$)} & \multicolumn{2}{c}{ Cooperation outside Team ($t_1$)} & \multicolumn{2}{c}{Prestige outside Team ($t_1$) }\\
\\ &  \multicolumn{4}{c|}{(in-degree)} & \multicolumn{2}{c}{(in-degree)} & \multicolumn{2}{c}{(in-degree) }\\
\\ 
[0.5ex]\cline{2-9}
 \hline \\ 
Exp. (Afinity-based groups)	&	1.94	&	3.97	&	1.15	&	5.05	&	1.94	&	3.97	&	1.94	&	3.97	\\	
Group Size (3)	&	1.52	&	1.54	&	1.51	&	1.51	&	1.52	&	1.54	&	1.52	&	1.54	\\	
Gender (Female)	&	1.02	&	1.02	&	1.04	&	1.04	&	1.02	&	1.02	&	1.02	&	1.02	\\	
Grade Level (10th)	&	1.71	&	1.71	&	1.65	&	1.66	&	1.71	&	1.71	&	1.71	&	1.71	\\	
Average grades	&	1.18	&	1.19	&	1.24	&	1.24	&	1.18	&	1.19	&	1.18	&	1.19	\\	
Team Friendship ($t_0$)	&	1.95	&	5.75	&		&		&	1.95	&	5.75	&	1.95	&	5.75	\\	
Exp.$*$ Team Friendship ($t_0$)	&		&	10.68	&		&		&		&	10.68	&		&	10.68	\\	
Friendship ($t_0$)	&		&		&	1.14	&	3.48	&		&		&		&		\\	
Exp.$*$ Friendship ($t_0$)	&		&		&		&	8.46	&		&		&		&		\\	
\hline
\end{tabular} 
}
\end{table}

\begin{table}[!htbp] \centering 
  \caption{Variance inflation factor (VIF) statistics of predictor and control variables in regressions for dependent variables related to social networks.} 
  \label{tab:s4} 
\begin{tabular}{lccc} 
\cline{1-4}\\
 & \multicolumn{3}{c}{\textit{Dependent variable:}} \\ 
[0.5ex]\cline{2-4}
\\ & \multicolumn{3}{c}{n$^{°}$ students with whom a person continues to work.} \\ 
[0.5ex]\cline{2-4}
	\\
 &	 Model (9)	&	 Model (10)	&	 Model (11)	
 \\
 \hline \\
Exp. (Afinity-based groups)	&	1.10	&	1.15	&	5.05	\\
Group Size (3)	            &	1.49	&	1.51	&	1.51	\\
Gender (Female)	            &	1.01	&	1.04	&	1.04	\\
Grade Level (10th)	        &	1.63	&	1.65	&	1.66	\\
Average grades	            &	1.17	&	1.24	&	1.24	\\
Friendship ($t_0$)	        &		    &	1.14	&	3.48	\\
Exp.$*$ Friendship ($t_0$)	&		    &		    &	8.46	\\
\\
\hline 
\end{tabular} 
\end{table}

\begin{table}[!htbp] \centering 
  \caption{Bootstapping regression coefficients for model Betweenness (model 1). Number of permutations = 2000} 
  \label{tab:s5} 
\begin{tabular}{lccc} 

\cline{1-4}
\\
Statistic&$\beta$&St. Dev & CI (95$\%$)\\
\\
\cline{1-4} 
 \\
Exp. (Afinity-based groups)	&	0.46	&	0.17	&[0.11	,	0.81]\\
Group Size (3)	&	-0.41	&	0.18	&[-0.78	,	-0.04]\\
Gender (Female)	&	0.08	&	0.12	&[-0.16	,	0.33]\\
Grade Level (10th)	&	0.08	&	0.15	&[-0.21	,	0.39]\\
Average grades	&	0.13	&	0.09	&[-0.05	,	0.32]\\
Team Friendship ($t_0$)	&	0.56	&	0.07	&[0.40	,	0.71]\\
Intercept	&	-0.95	&	0.13	&[-1.23	,	-0.67]\\
\hline 
\end{tabular} 
\end{table}

\begin{table}[!htbp] \centering 
  \caption{Bootstapping regression coefficients for model Betweenness (model 2). Number of permutations = 2000} 
  \label{tab:s6} 
\begin{tabular}{lccc} 

\cline{1-4}
\\
Statistic&$\beta$&St. Dev & CI (95$\%$)\\
\\
\cline{1-4} 
 \\
Exp. (Afinity-based groups)	&	1.15	&	0.24	&[0.68	,	1.62]\\
Group Size (3)	&	-0.35	&	0.17	&[-0.70	,	-0.01]\\
Gender (Female)	&	0.13	&	0.11	&[-0.10	,	0.36]\\
Grade Level (10th)	&	0.08	&	0.14	&[-0.20	,	0.36]\\
Average grades	&	0.20	&	0.09	&[0.02	,	0.38]\\
Team Friendship ($t_0$)	&	1.02	&	0.14	&[0.75	,	1.29]\\
Exp.$*$ Team Friendship ($t_0$)	&	-0.63	&	0.16	&[-0.95	,	-0.32]\\
Intercept	&	-1.30	&	0.15	&[-1.61	,	-0.99]\\

\hline 
\end{tabular} 
\end{table}

\begin{table}[!htbp] \centering 
  \caption{Bootstapping regression coefficients for model Betweenness (model 3). Number of permutations = 2000} 
  \label{tab:s7} 
\begin{tabular}{lccc} 

\cline{1-4}
\\
Statistic&$\beta$&St. Dev & CI (95$\%$)\\
\\
\cline{1-4} 
 \\
Exp. (Afinity-based groups)	&	1.05	&	0.18	&[0.69	,	1.39]\\
Group Size (3)	&	-0.74	&	0.22	&[-1.17	,	-0.31]\\
Gender (Female)	&	-0.15	&	0.15	&[-0.45	,	0.16]\\
Grade Level (10th)	&	0.55	&	0.19	&[0.17	,	0.93]\\
Average grades	&	0.05	&	0.12	&[-0.19	,	0.28]\\
Friendship ($t_0$)	&	0.09	&	0.03	&[0.03	,	0.14]\\
Intercept	&	-0.96	&	0.20	&[-1.35	,	-0.57]\\

\hline 
\end{tabular} 
\end{table}

\begin{table}[!htbp] \centering 
  \caption{Bootstapping regression coefficients for model Betweenness (model 4). Number of permutations = 2000} 
  \label{tab:s8} 
\begin{tabular}{lccc} 

\cline{1-4}
\\
Statistic&$\beta$&St. Dev & CI (95$\%$)\\
\\
\cline{1-4} 
 \\
Exp. (Afinity-based groups)	&	1.10	&	0.34	&[0.42	,	1.77]\\
Group Size (3)	&	-0.74	&	0.22	&[-1.17	,	-0.31]\\
Gender (Female)	&	-0.15	&	0.15	&[-0.45	,	0.16]\\
Grade Level (10th)	&	0.55	&	0.19	&[0.17	,	0.93]\\
Average grades	&	0.04	&	0.12	&[-0.19	,	0.28]\\
Friendship ($t_0$)	&	0.09	&	0.06	&[-0.02	,	0.20]\\
Exp.$*$ Friendship ($t_0$)	&	-0.01	&	0.06	&[-0.13	,	0.11]\\
Intercept	&	-1.00	&	0.29	&[-1.58 ,	-0.42]\\

\hline 
\end{tabular} 
\end{table}

\begin{table}[!htbp] \centering 
  \caption{Bootstapping regression coefficients for model Betweenness (model 5). Number of permutations = 2000} 
  \label{tab:s9} 
\begin{tabular}{lccc} 

\cline{1-4}
\\
Statistic&$\beta$&St. Dev & CI (95$\%$)\\
\\
\cline{1-4} 
 \\

Exp. (Afinity-based groups)	&	-0.71	&	0.21	&[-1.13	,	-0.29]\\
Group Size (3)	&	-1.28	&	0.23	&[-1.73	,	-0.83]\\
Gender (Female)	&	-0.25	&	0.15	&[-0.55	,	0.05]\\
Grade Level (10th)	&	1.06	&	0.19	&[0.68	,	1.42]\\
Average grades	&	0.04	&	0.12	&[-0.19	,	0.27]\\
Team Friendship ($t_0$)	&	0.16	&	0.10	&[-0.03	,	0.35]\\
Intercept	&	0.06	&	0.17	&[-0.28	,	0.39]\\

\hline 
\end{tabular} 
\end{table}

\begin{table}[!htbp] \centering 
  \caption{Bootstapping regression coefficients for model Betweenness (model 6). Number of permutations = 2000} 
  \label{tab:s10} 
\begin{tabular}{lccc} 
\cline{1-4} 
 \\
Statistic&$\beta$&St. Dev & CI (95$\%$)\\
\\
\cline{1-4} 
 \\
Exp. (Afinity-based groups)	&	-1.16	&	0.31	&[-1.77	,	-0.55]\\
Group Size (3)	&	-1.32	&	0.22	&[-1.76	,	-0.87]\\
Gender (Female)	&	-0.28	&	0.15	&[-0.58	,	0.02]\\
Grade Level (10th)	&	1.06	&	0.18	&[0.70	,	1.43]\\
Average grades	&	-0.00	&	0.12	&[-0.23	,	0.23]\\
Team Friendship ($t_0$)	&	-0.14	&	0.18	&[-0.49	,	0.21]\\
Exp.$*$ Team Friendship ($t_0$)	&	0.41	&	0.20	&[0.00	,	0.82]\\
Intercept	&	0.28	&	0.20	&[-0.12	,	0.68]\\

\\
\cline{1-4}

\hline 
\end{tabular} 
\end{table}

\begin{table}[!htbp] \centering 
  \caption{Bootstapping regression coefficients for model Betweenness (model 7). Number of permutations = 2000} 
  \label{tab:s11} 
\begin{tabular}{lccc} 

\cline{1-4}
\\
Statistic&$\beta$&St. Dev & CI (95$\%$)\\
\\
\cline{1-4} 
 \\
Exp. (Afinity-based groups) & -0.22 & 0.26 & [-0.73 , 0.29]\\
Group Size (3)              & -0.24 & 0.27 & [-0.79 , 0.30]\\
Gender (Female)             & -0.14 & 0.18 & [-0.51 , 0.22]\\
Grade Level (10th)          & -0.03 & 0.23 & [-0.47 , 0.42]\\
Average grades              & 0.10  & 0.14 & [0.01 , 0.58]\\
Team Friendship ($t_0$)     & 0.55  & 0.11 & [0.32 , 0.77]\\
Intercept                   & -0.66 & 0.20 & [-1.06 , -0.25]\\

\hline 
\end{tabular} 
\end{table}

\begin{table}[!htbp] \centering 
  \caption{Bootstapping regression coefficients for model Betweenness (model 8). Number of permutations = 2000} 
  \label{tab:s12} 
\begin{tabular}{lccc} 

\cline{1-4}
\\
Statistic&$\beta$&St. Dev & CI (95$\%$)\\
\\
\cline{1-4} 
 \\
Exp. (Afinity-based groups)      & -0.07 & 0.38 & [-0.82 , 0.68]\\
Group Size (3)                   & -0.23 & 0.28 & [-0.78 , 0.32]\\
Gender (Female)                  & -0.13 & 0.18 & [-0.50 , 0.23]\\
Grade Level (10th)               & -0.03 & 0.23 & [-0.48 , 0.42]\\
Average grades                   & 0.04  & 0.14 & [0.01 , 0.40]\\
Team Friendship ($t_0$)          & 0.65  & 0.22 & [0.21 , 1.08]\\
Exp.$*$ Team Friendship ($t_0$)  & -0.14 & 0.25 & [-0.64 , 0.36]\\
Intercept                        & -0.73 & 0.25 & [-1.22 , -0.24]\\

\hline 
\end{tabular} 
\end{table}

\begin{table}[!htbp] \centering 
  \caption{Bootstapping regression coefficients for model Betweenness (model 9). Number of permutations = 2000} 
  \label{tab:s13} 
\begin{tabular}{lccc} 

\cline{1-4}
\\
Statistic&$\beta$&St. Dev & CI (95$\%$)\\
\\
\cline{1-4} 
 \\
Exp. (Afinity-based groups)	&	1.87	&	0.13	&	[1.60 , 2.15]	\\
Group Size (3)	&	-0.37	&	0.18	&	[-0.74 , -0.01]	\\
Gender (Female)	&	-0.10	&	0.12	&	[-0.35 , 0.13]	\\
Grade Level (10th)	&	0.51	&	0.15	&	[0.21 , 0.81]	\\
Average grades	&	0.25	&	0.09	&	[0.01 , 0.44]	\\
Intercept	&	0.20	&	0.12	&	[-0.05 , 0.46]	\\

\\
\hline 
\end{tabular} 
\end{table}

\begin{table}[!htbp] \centering 
  \caption{Bootstapping regression coefficients for model Betweenness (model 10). Number of permutations = 2000} 
  \label{tab:s14} 
\begin{tabular}{lccc} 

\cline{1-4}
\\
Statistic&$\beta$&St. Dev & CI (95$\%$)\\
\\
\cline{1-4} 
 \\
Exp. (Afinity-based groups)	&	1.90	&	0.14	&	[1.60	,	2.19]	\\
Group Size (3)	&	-0.37	&	0.18	&	[-0.74	,	-0.01]	\\
Gender (Female)	&	-0.09	&	0.12	&	[-0.35	,	0.16]	\\
Grade Level (10th)	&	0.49	&	0.16	&	[0.17	,	0.81]	\\
Average grades	&	0.26	&	0.09	&	[0.01	,	0.46]	\\
Friendship ($t_0$)	&	-0.01	&	0.02	&	[-0.05	,	0.03]	\\
Intercept	&	0.24	&	0.16	&	[-0.08	,	0.57]	\\
\\
\hline 
\end{tabular} 
\end{table}

\begin{table}[!htbp] \centering 
  \caption{Bootstapping regression coefficients for model Betweenness (model 11). Number of permutations = 2000} 
  \label{tab:s15} 
\begin{tabular}{lccc} 

\cline{1-4}
\\
Statistic&$\beta$&St. Dev & CI (95$\%$)\\
\\
\cline{1-4} 
 \\
Exp. (Afinity-based groups)	&	1.60	&	0.28	&	[1.04	,	2.17]	\\
Group Size (3)	&	-0.38	&	0.18	&	[-0.74	,	-0.02]	\\
Gender (Female)	&	-0.09	&	0.12	&	[-0.35	,	0.16]	\\
Grade Level (10th)	&	0.49	&	0.16	&	[0.17	,	0.81]	\\
Average grades	&	0.17	&	0.09	&	[0.01	,	0.37]	\\
Friendship ($t_0$)	&	-0.05	&	0.04	&	[-0.15	,	0.03]	\\
Exp.$*$ Friendship ($t_0$)	&	0.06	&	0.05	&	[-0.03	,	0.16]	\\
Intercept	&	0.46	&	0.24	&	[-0.02	,	0.95]	\\

\hline 
\end{tabular} 
\end{table}

\end{document}